\documentclass[twocolumn,aps,prd,tighten,amssymb,,amsmath,showpacs]{revtex4}
\usepackage{amssymb}
\usepackage{amsmath}
\usepackage{amsthm}
\usepackage{epsfig}

\begin{document}

\title{Multi-state Boson Stars}

\date{\today}
\label{firstpage}

\author{A.~Bernal$^{1}$, J.~Barranco$^{1}$, D.~Alic$^{1,2}$ 
        and C.~Palenzuela$^{1,3}$}
\affiliation{
$^{1}$Max-Planck-Institut f\"ur Gravitationsphysik, Albert Einstein Institut, 14476 Golm, Germany. \\
$^{2}$Department of Physics, Universitat de les Illes Balears, Cra. Valldemossa, Spain. \\
$^{3}$Canadian Institute for Theoretical Astrophysics (CITA), Toronto, Canada. \\}

\begin{abstract}
Motivated by the increasing interest in models which consider scalar
fields as viable dark matter candidates, we have
constructed a generalization of relativistic Boson Stars (BS) composed of two 
coexisting states of the scalar field, the ground state and the first excited state. 
We have studied the dynamical evolution of these Multi-state Boson Stars
(MSBS) under radial perturbations, using numerical techniques.
We show that stable MSBS can be constructed, when the number of particles in the first excited 
state, $N^{(2)}$, is smaller than the number of particles in the ground state, $N^{(1)}$. 
On the other hand, when $N^{(2)} > N^{(1)}$, the configurations are initially
unstable. However, they evolve and settle down into stable configurations. 
In the stabilization process, the initially ground state is excited and ends in a first 
excited state, whereas the initially first excited state ends in a ground state. During this 
process, both states emit scalar field radiation, decreasing their number of particles. 
This behavior shows that even though BS in the first excited state are intrinsically unstable 
under finite perturbations, the configuration resulting from the combination of this state with 
the ground state produces stable objects.  
Finally we show in a qualitative way, that stable MSBS could be realistic models of 
dark matter galactic halos, as they produce rotation curves that are flatter at large radii
than the rotation curves produced by BS with only one state.
\end{abstract}
\maketitle


\section{Introduction}
\label{introduction}

The existence of Dark Matter (DM) in the Universe is strongly supported by astronomical 
observations that range from galactic up to cosmological scales
(see for example \cite{Sahni:2004ai} and references therein). 
Observations indicate that stars rotate too fast around the center of the galaxy to be 
bound by Newtonian gravity if all matter is visible \cite{1978ApJ...225L.107R,Rubin:1980zd}.
This issue, known as the Rotation Curves (RC) problem, implies within the context of Einstein's 
General Relativity, that a great amount of the matter in the galaxy is invisible.
The nature of this dark matter, that has a negligible interaction with the visible
matter and whose presence is only observed trough its gravitational effects,  
is still unknown. The most popular candidate are the so-called weakly interacting 
massive particles (WIMPs) \cite{Freedman:2003ys,Jungman:1995df}, 
leading to the standard Cold Dark Matter model.
This scenario is very successful at a cosmological level, as its predictions are in 
good agreement with the observational data \cite{DelPopolo:2008mr,Samtleben:2007zz}. However, it has difficulties in fitting 
the observations at a galactic level \cite{Navarro:1996gj,Moore:1999gc,McGaugh:2001yc,Moore:1994yx,1999ApJ...522...82K}. 
If DM is modeled by WIMPs, one obtains a cuspy density profile of the DM in the galaxy.
But high resolution data of low surface brightness galaxies, which are composed mainly of DM, imply that their DM distribution has a flat core \cite{McGaugh:2001yc,Moore:1994yx}. 
This model also fails in predicting the number of satellite galaxies around each galactic 
halo, exceeding far beyond what is observed around the Milky Way \cite{1999ApJ...522...82K}.

A different approach consists in describing the dark matter as a scalar field \cite{Sahni:1999qe,Matos:2000ng,Hu:2000ke}. The Scalar Field Dark Matter (SFDM) model 
has been proved to be successful at cosmological scales \cite{UrenaLopez:2000aj}.
This model can also avoid the problems that WIMPs present at a galactic level,
producing a non-cuspy density profile \cite{Peebles:2000yy,Matos:2003pe,UrenaLopez:2000aj} and explaining the dearth of satellite galaxies around each galactic halo \cite{UrenaLopez:2000aj}. Because of the viability presented by the SFDM model, it is stimulating to go further on testing it. 
For instance, the model has to reproduce the observed RC of galaxies.
At this point Boson-Star-like objects could play an important role.

In the SFDM model, the dark matter particle is an 
ultra-light massive spinless boson ($m\sim 10^{-23}$eV \cite{UrenaLopez:2000aj}). 
These bosons could collapse forming gravitationally 
bounded structures. With such ultra-light mass, the boson's Compton wave length 
is of the order of kilo-parsecs, and structures with 
comparable length scales - like galactic halos - could be formed
as condensates described by a coherent scalar field \cite{Matos:2008ag}.
These condensates can be associated with Boson stars (BS), which are solutions 
of the Einstein-Klein-Gordon equations where the gravity attraction is
balanced by the dispersive character of the scalar field \cite{Schunck:2003kk,Jetzer:1991jr}. 
BS were first studied by Kaup 
\cite{PhysRev.172.1331} and a year later by
Ruffini and Bonazzola \cite{PhysRev.187.1767}, who settled two different treatments. The first one,
developed in \cite{PhysRev.172.1331}, is a completely classical treatment with
a massive complex scalar field minimally coupled to gravity. 
In the second one \cite{PhysRev.187.1767}, a real quantized scalar field is introduced in order to
describe a many boson system though
maintaining the geometry as a classical entity (i.e., a semiclassical limit is adopted).
The relevant quantity computed in this case is the mean value 
$\langle Q|\hat T^{\mu \nu}|Q\rangle$ of the energy momentum tensor operator, 
where $|Q\rangle$ is the state of the system of many particles. 
When one considers a $|Q\rangle$ for which all the particles are in the same state, it turns out that the mean value of $\hat T_{\mu\nu}$ generates the same energy-momentum 
tensor as the complex classical field and consequently, the same macroscopic results, i.e. both a quantized real scalar field and a classical complex scalar field yield to the same self-gravitating system.

Until very recently, only boson stars with all the particles in one state have been
considered. However, Newtonian configurations with scalar fields coexisting in the ground 
and excited states were introduced in \cite{Matos:2007zza} in order to model
dark matter halos. Previous studies in modeling dark matter halos using newtonian BS were
done in \cite{1994PhRvD..50.3650S,Ji:1994xh,Lee:1995af,Guzman:2006yc,Arbey:2003sj}.
However, these structures can not account for a realistic halo since the configurations
in the ground state produce RC which are not flat enough at large radii.
The case of a massless scalar field used in order to
fit rotation curve data of several galaxies was considered in 
\cite{Schunck:1998nq}. Nevertheless, it has been proved in 
\cite{Goldwirth:1987nu,Seidel:1993zk} that no nonsingular self-gravitating
solitonic objects can be formed with a massless scalar field.
On the other hand, RC from excited BS are in better agreement with the astrophysical observations, 
but these structures are unstable \cite{1998PhRvD..58j4004B}.

In \cite{arge-luis} it was shown that Newtonian mixed configurations 
could account for more realistic 
DM halos, as they are stable and could fit with better agreement the observed RCs
even at large radii.
In the present work, we are interested in the fully relativistic generalization of these
Multi State Boson Stars (MSBS).
The idea is to consider the possibility that the bosons are not all in the same state,
but rather populating different coexisting states, as was already pointed out in \cite{PhysRev.187.1767}. 
It turns out (see \cite{PhysRev.187.1767} and Appendix  \ref{quantized}) 
that the resulting equations for a MSBS in the semiclassical approach are 
equivalent to the case where 
a collection of complex classical scalar fields is considered,
one for each state, which are only coupled through gravity. 
Without any loss of generality, we can choose either the semiclassical formulation 
of such MSBS or its pure classical counterpart. 
We will follow the latter approach in order to investigate MSBS,
analyzing in detail the properties and stability of configurations with two states, 
a ground state and a first excited state. 

The previous stability studies of BS can be divided roughly in two categories, depending
on the type of perturbations considered:
\begin{enumerate}
\item  Studies where the perturbations preserve the number of particles (infinitesimal
perturbations), which generally involve linear perturbation analysis 
\cite{1988PhRvD..38.2376G,1989NuPhB.315..477L} and catastrophe theory
\cite{Kusmartsev:1990cr}.
\item Those where the perturbations  don't conserve the number of particles 
(finite perturbations), which have been addressed mainly by numerical studies.
\cite{1990PhRvD..42..384S, 1998PhRvD..58j4004B,Hawley:2000dt}. Furthermore,
the late time evolution of unstable boson star under finite perturbation can only
be followed by numerical simulations.
\end{enumerate}
A consistent result coming from both type of studies is that BS in the ground state are stable against 
perturbations if the amplitude of the scalar field at the origin $\phi(0)$ is smaller than 
the critical value $\phi_{max}(0)$ where the maximum mass $M_{max}$ is reached. In the case of 
excited BS there are some important differences. Although the linear stability analysis
shows stability up to the critical value $\phi_{max}(0)$ \cite{1989NuPhB.315..477L,Jetzer:1991jr}
when the number of particles is not conserved, excited BS are intrinsically unstable 
even for $\phi(0) \le \phi_{max}(0)$, since
finite perturbations drive the star either to collapse to a black hole or to 
decay to the ground state \cite{1989NuPhB.315..477L,1998PhRvD..58j4004B}.

From these results one could infer that the MSBS states would be unstable under perturbations when
the number of particles is allowed to change, since they contain at least one excited state. 
Quite surprisingly, our numerical analysis
shows that there is a region of the solution space with stable configurations. Roughly
speaking, the ground state produces a deeper gravitational potential which can be enough
to stabilize the excited state. 

This paper is organized as follows. In Section \ref{equations}, we present the
formalism used for the numerical evolution of the Einstein-Klein-Gordon 
system, describing a MSBS in the classical approximation.
Section \ref{initial_data} describes how the initial data for a MSBS with two different
states is constructed. In Section \ref{stability}, we present numerical results obtained
from the evolution of the two-state boson stars. We study two features of the evolution,
namely the stability and the late time behavior.
In both cases, we add an small perturbation. 
In the first case, the perturbed MSBS is evolved for short time scales, in order to study the 
behavior of the perturbations. In the second case, we evolve unstable MSBS for longer times following the
properties of the resulting configurations.
We compute in section \ref{DMhalos} RC from stable MSBS and discuss qualitatively why 
these RC are in better agreement with the observed RC of galaxies.
We conclude in Section~\ref{conclusions}.
The description of the semiclassical approach 
is presented in Appendix~\ref{quantized}, while 
Appendix~\ref{the_Z3_system} is devoted to a detailed description of the evolution
equations used for numerical evolution.


\section{The Einstein-Klein-Gordon system}
\label{equations}

Let us consider a semi-classical real massive scalar field with $P$ different
excited states, which is equivalent to considering a collection of $P$
classical complex scalar fields (one for each state) coupled only through gravity.
In a curved spacetime, the dynamics of these
MSBS can be described by the following Lagrangian density (adopting geometrical units, i.e. $G=c=\hbar=1$),
\begin{eqnarray}\label{Lagrangian}
  {\cal L} = - \frac{1}{16 \pi} R
   + \sum_{n=1}^{P} \frac{1}{2} \left[ g^{ab} \partial_a \bar \phi^{(n)} \partial_b \phi^{(n)}
   + V\left( \left|\phi^{(n)} \right|^2\right) \right]
\end{eqnarray}
where $R$ is the Ricci scalar, $g_{ab}$ is the spacetime metric,
$\phi^{(n)}$ are the scalar fields,  $\bar \phi^{(n)}$ their complex conjugate,
and $V(|\phi^{(n)}|^2)$ a  potential depending only on $|\phi^{(n)}|^2$.
Throughout this paper, Roman letters from the beginning of the
alphabet $a,b,c,..$ denote spacetime indices ranging from 0 to 3,
while letters near the middle $i,j,k,..$ range from 1 to 3,
denoting spatial indices.
This Lagrangian gives rise to the equations determining the
evolution of the metric (Einstein equations) and those governing
the scalar fields behavior (Klein-Gordon equations).



The variation of the action associated with the Lagrangian (\ref{Lagrangian})
with respect to the metric $g_{ab}$, leads to the well-known Einstein equations 
\begin{eqnarray}\label{EE1}
   R_{ab} - \frac{R}{2} g_{ab} = 8 \pi T_{ab},
\end{eqnarray}
where $R_{ab}$ is the Ricci tensor. $T_{ab}$ is
the total stress-energy tensor, given by the addition of the single
stress-energy tensors of each scalar field, namely
\begin{eqnarray}\label{stress-energy}
   T_{ab} &=& \sum_{n=1}^{P} {T_{ab}}^{(n)}, \\
   {T_{ab}}^{(n)}  &=& \frac{1}{2} \left[\partial_a \bar \phi^{(n)} \partial_b \phi^{(n)}
                        + \partial_a \phi^{(n)}~ \partial_b \bar \phi^{(n)} \right]
   \nonumber \\
    &-& \frac{1}{2} g_{ab} \left[g^{cd}\partial_c \bar \phi^{(n)}  \partial_d \phi^{(n)}
        + V\left(|\phi^{(n)}|^2\right)\right].
\end{eqnarray}
The Einstein equations form a system of $10$ non-linear partial
differential equations for the spacetime metric components
$g_{ab}$. 



On the other hand, the variation of the Lagrangian (\ref{Lagrangian})
with respect to each scalar field $\phi^{(n)}$, leads to a set of
Klein-Gordon (KG) equations which are only coupled through the gravity,
\begin{equation}\label{KG1}
  \Box \phi^{(n)} = \frac{d V}{d |\phi^{(n)}|^2} \phi^{(n)} \, ,
\end{equation}
where the box $\Box = g^{ab} \nabla_a \nabla_b$ stands for
the wave operator on a curved background. In the following, we will restrict
ourselves to the free field case, where the potential takes the form
\begin{equation}\label{potential}
  V( |\phi^{(n)}|^2 ) = m^2~ |\phi^{(n)}|^2 \, ,
\end{equation}
with $m$ a parameter that can be identified with the bare
mass of the field theory.

The matter Lagrangian is invariant under global U(1) transformations
\begin{equation}
         \phi^{(n)} \rightarrow \phi^{(n)} e^{i \varphi^{(n)}}.
\end{equation}
This symmetry implies that there is a set of Noether currents densities
$J^{(n)}_{a}$,
\begin{equation}
  J^{(n)}_{a} = \frac{i}{2} \sqrt{-g} \left[\bar \phi^{(n)} ~\partial_{a} \phi^{(n)}-
  \phi^{(n)} ~\partial_{a} \bar \phi^{(n)} \right],
\end{equation}
satisfying for each $n$ the conservation law $\nabla^a J^{(n)}_{a} = 0$. The
Noether charge contained in some radius is given by
\begin{equation}\label{Number_particles_radius}
   N^{(n)}(r)  = \int_0^r g^{0a} ~J^{(n)}_{a}~ dx^3~,
\end{equation}
so that the total Noether charge of the system $N$ is the sum of the total
individual ones $N^{(n)} \equiv N^{(n)}(\infty)$, namely
\begin{equation}\label{Number_particles}
   N = \sum_{n=1}^{P} N^{(n)} ~~.
\end{equation}
As discussed in ~\cite{PhysRev.187.1767}, this quantity $N$ can be associated 
with the total number of bosonic particles. Consequently,
$N^{(n)}$ can be interpreted as the number of particles in the state labeled
by $n$.


\section{Initial Data for Multi-State Boson Stars}
\label{initial_data}

The initial data for the MSBS configurations is computed 
in spherical symmetry with a one-dimensional code. We adopt the following
harmonic ansatz for each scalar field,
\begin{eqnarray}
\label{oscillate_phi}
\phi^{(n)}(t,r) &=& \phi_n(r) \,e^{-i \omega_n t}.
\end{eqnarray}
With this assumption, the source for the Einstein equations becomes time
independent. Our goal is to find $\{\phi_n(r),\omega_n\}$ and the
metric coefficients, such that the spacetime generated
by this matter configuration is static. 

We begin by considering the problem in
polar-areal coordinates \cite{PhysRev.172.1331,
2005PhDT.........2L}. The line element in these coordinates takes the form
\begin{equation}\label{oscillate_phi_metric}
ds^2 = - \alpha\left(r\right)^2 dt^2 + a\left(r\right)^2 dr^2 + r^2 d\Omega^2.
\end{equation}
Then the equilibrium equations, obtained by substituting the
ansatz (\ref{oscillate_phi}) and the metric (\ref{oscillate_phi_metric})
in the Einstein-Klein-Gordon system (\ref{EE1},\ref{KG1}), are given by
\begin{eqnarray}
\nonumber
\label{ekg1}
\partial_r a      &=& \frac{a}{2}      \left\{-    \frac{a^2-1}{r} \right.\\
  && \left. + 4 \pi r \sum_{n=1}^{P} \left[ \left(\frac{\omega_n^2}{\alpha^2} + m^2\right)
               a^2 \phi_n^2  + \Phi^2_n \right]  \right\}, \\ 
\nonumber
\partial_r \alpha &=&  \frac{\alpha}{2}\left\{\ \ \frac{a^2-1}{r} \right.\\
   && \left. + 4 \pi r \sum_{n=1}^{P} \left[ \left(\frac{\omega_n^2}{\alpha^2}-m^2\right)
               a^2 \phi_n^2 + \Phi^2_n \right] \right\},
\label{slicing} \\
\partial_r \phi_n &=& \Phi_n,\\
\nonumber
\partial_r \Phi_n &=& - \left\{ 1 + a^2 - 4 \pi r^2 a^2 m^2 
              \left( \sum_{s=1}^{P} \phi_s^2 \right) \right\} \frac{\Phi_n}{r} \\
               && - \left( \frac{\omega_n^2}{\alpha^2} - m^2 \right) \phi_n\, a^2.\label{ekg2}
\end{eqnarray}

In order to obtain a solution of this system, we provide the following
boundary conditions, motivated by the physical
situation under study,
\begin{eqnarray}
\phi_n\left(0 \right) &=& \phi_{cn}, \label{BC1}\\
\Phi_n\left(0 \right) &=& 0,\label{BC2}\\
a\left(0 \right)      &=& 1,\label{BC4} \\
\lim_{r\rightarrow \infty}\phi_n\left(r \right) &\approx& 0, \label{BC5}\\
\lim_{r\rightarrow \infty}\alpha\left(r \right) &=& \lim_{r\rightarrow \infty} \frac{1}{a(r)} \label{BC6}~,
\end{eqnarray}
which guarantee regularity at the origin and asymptotic flatness.
For given central values of the fields $\{\phi_{cn}\}$, 
we only need to adjust the eigenvalues $\{\omega_n\}$ and the value  $\alpha(0)$ in order to generate a solution with the appropriate asymptotic behavior (\ref{BC5}-\ref{BC6}). This is a  shooting problem that we solve by integrating from $r=0$ towards the outer boundary $r=r_{out}$, with a second order shooting method. The boundary conditions for the scalar fields at $r_{out}$ are imposed considering that localized solutions decrease asymptotically as $\phi_n\sim \exp \left( -\sqrt{m^2-\omega_n^2}r\right)/r$ in a 
Schwarzschild-type asymptotic background. At the outer boundary, the conditions
are
\begin{eqnarray}\label{BC7}
\phi_n\left(r_{out} \right)\left(\sqrt{m^2-\omega_n^2}+\frac{1}{r_{out}^2} \right)+ \Phi_n\left(r_{out} \right)= 0.
\end{eqnarray}
The shooting procedure is performed for different values of $r_{out}$. As $r_{out}$
is increasing, the shooting parameters converge, and we choose the solution as the one which 
satisfies the conditions (\ref{BC7}) for some $r_{out}$ within a prescribed tolerance. From this point on, we match to the scalar fields and the metric coefficients their asymptotic behavior.

A qualitative characteristic of the radial functions $\phi_n$ is their number of nodes (ie,
how many times they do cross zero), which reflects the excited state of the boson star.
If the radial function does not have any node, the boson star is in the ground state.
When there is a node, the boson star is in the first excited state, and so on. In the next
subsection we construct initial configurations with two scalar fields $P=2$,
one in the ground state and the other in the first excited state. Notice that this is the simplest 
non-trivial configuration, since the MSBS with two scalar fields in the ground state can be reduced 
to one scalar field solution by redefining the scalar fields. This is a consequence of 
the indistinguishably of the boson particles in the same state.

Once the solution is computed in this coordinate system, a change of coordinates is performed to 
maximal isotropic ones,
\begin{equation}
ds^2 = \alpha^2\left(\tilde r \right) dt^2 + \psi^4\left(\tilde r \right) 
\left( d\tilde r^2 + \tilde r^2 d\Omega^2 \right),
\end{equation}
which are more convenient for our numerical evolutions.
Finally, a simple inspection on the system (\ref{ekg1}-\ref{ekg2}) shows that the re-definition
\begin{equation}\label{units}
\tilde r = r m\,, \quad \tilde \omega_n = \frac{\omega_n}{m}\,,
\end{equation} 
leads to a set of equations that no longer have $m$ on it. 
Hence, the selection of geometrical units (e.g. $G=\hbar=c=1$) and the
redefinition (\ref{units}) give us dimensionless units for $\tilde r, \tilde \omega_n$,
which is equivalent to choosing $m=1$ in our equations.
Throughout this paper, we use these dimensionaless coordinates.


\subsection{Configurations of ground and excited states}

\begin{figure}[h]
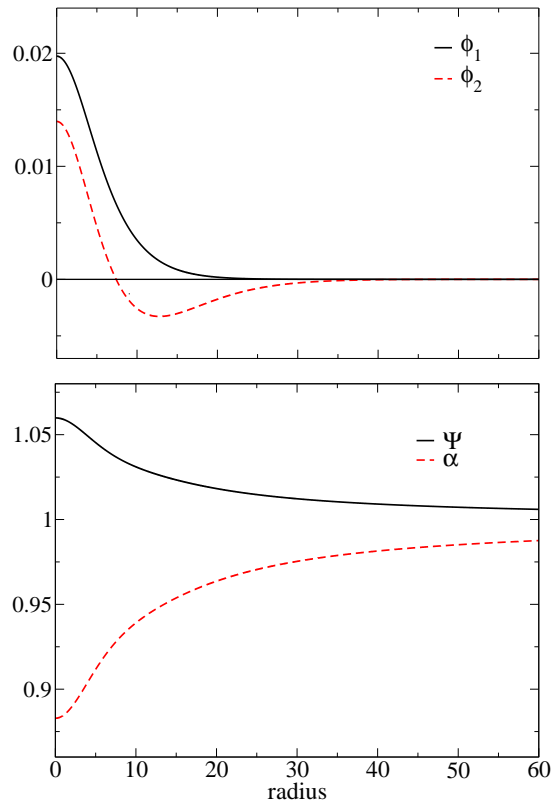

\includegraphics[angle=0,width=0.4\textwidth]{mixedID.eps}
\includegraphics[angle=0,width=0.4\textwidth]{mixedID2.eps}
\caption{Ground-1st excited configuration for $\phi_1(0)=0.0197$ and 
fraction $\eta=1$. The upper panel corresponds to the initial profiles of the two scalar fields,
and the lower panel, to the lapse function $\alpha$ and the conformal
factor $\Psi$.}
\label{mixedID}
\end{figure}

Let us consider the simplest non-trivial case with only two scalar fields $P=2$,
one with $N^{(1)}$ particles in the ground state, and the other with $N^{(2)}$ particles
in the first excited state. A useful way to construct the initial data is specifying the
fraction between the number of particles in each state of the configuration,
\begin{equation}
 \eta= \frac{N^{(2)}}{N^{(1)}}.
\end{equation}
In this case, we complete the system (\ref{ekg1}-\ref{ekg2}) with the differential expressions 
for the number of particles in each state
\begin{equation}\label{dNdr}
    \partial_r N^{(n)}(r)=4\pi\frac{a}{\alpha} \omega_n \phi_n^2 r^2,
\end{equation}
with boundary conditions, $N^{(n)}(0)=0$. If 
$\eta$ is specified, it is sufficient to prescribe as boundary conditions the central value of only one of the scalar fields, for instance, $\phi_{c1}$. The new system of equations (\ref{BC7}) 
and (\ref{dNdr}) becomes a shooting problem for the four parameters 
$\{\omega_1,\omega_2,\alpha(0),\phi_2(0)\}$. For a specific fraction $\eta$,
it is necessary to adjust the four parameters such that equation (\ref{BC7}) and the condition $N^{(2)}(r_{out})=\eta N^{(1)}(r_{out})$ are satisfied. 

Fig.~\ref{mixedID} shows an example of the radial profiles of the two scalar fields, lapse and conformal factor, for a MSBS with $\{\eta=1,\phi_1(0)=0.0197\}$.

Two important characteristics of MSBS are the total gravitational mass $M$ and the
radius $R_{99}$. The first one is calculated as
\begin{equation}\label{massMSBS}
M=\frac{r_{out}}{2}\left(1-\frac{1}{a^2(r_{out})}\right)\,,
\end{equation}
and the radius $R_{99}$ is defined as the radius where $M$ reaches the $99\%$ of its value.  
The choice of geometrical units, plus the re-definition of the
coordinates eq. (\ref{units}), imply that both $R_{99}$ and $M$ are dimensionless 
variables. The physical units can be recovered by using the following relations:
\begin{equation}\label{unidades}
M^{Physical}=M \frac{m_p^2}{m} \qquad R_{99}^{Physical}=R_{99} \frac{\hbar}{m c} 
\end{equation} 
where $m_p$ the Planck's mass and $m$ the mass of the boson associated to the scalar field.

\begin{figure}
\includegraphics[width=0.23\textwidth]{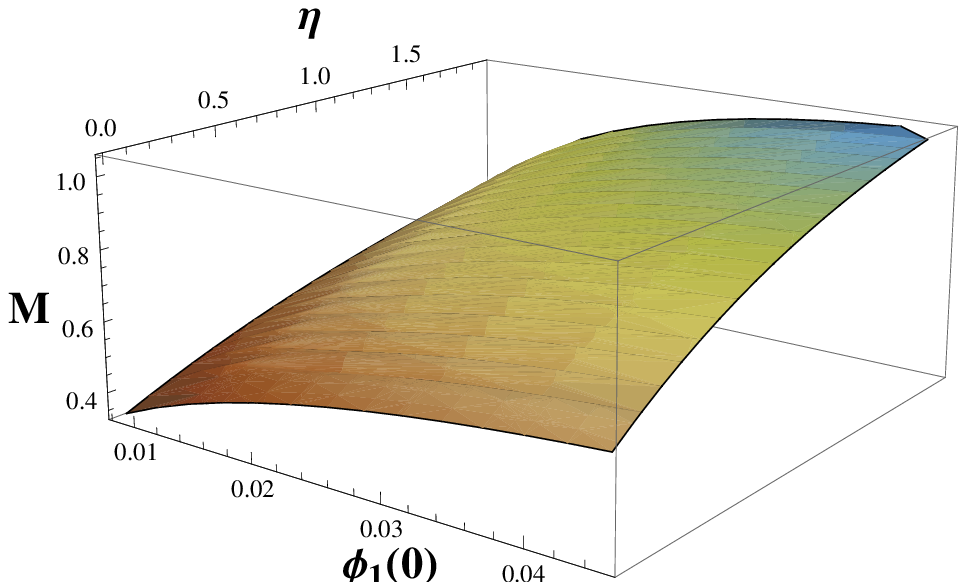}
\includegraphics[width=0.23\textwidth]{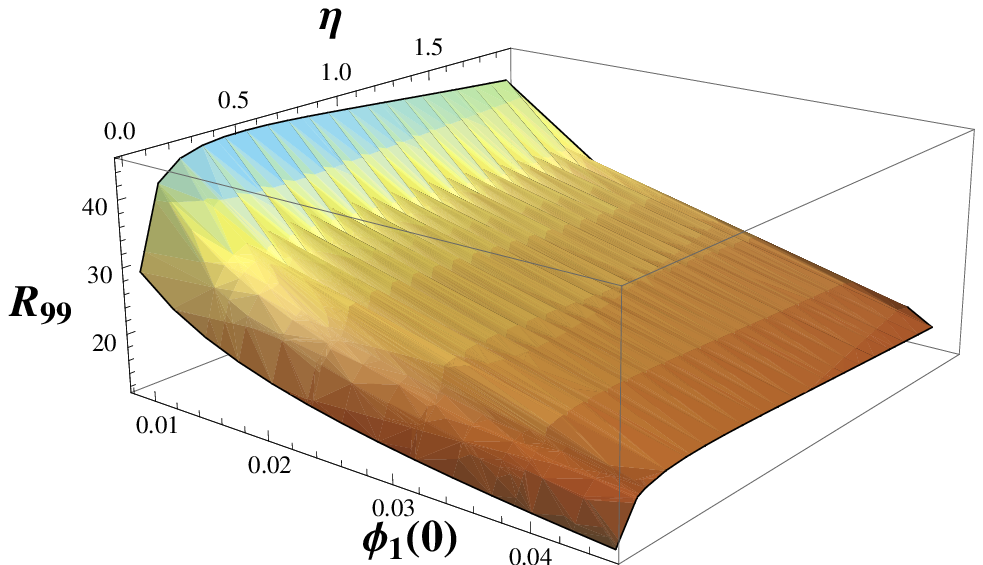}
\includegraphics[width=0.23\textwidth]{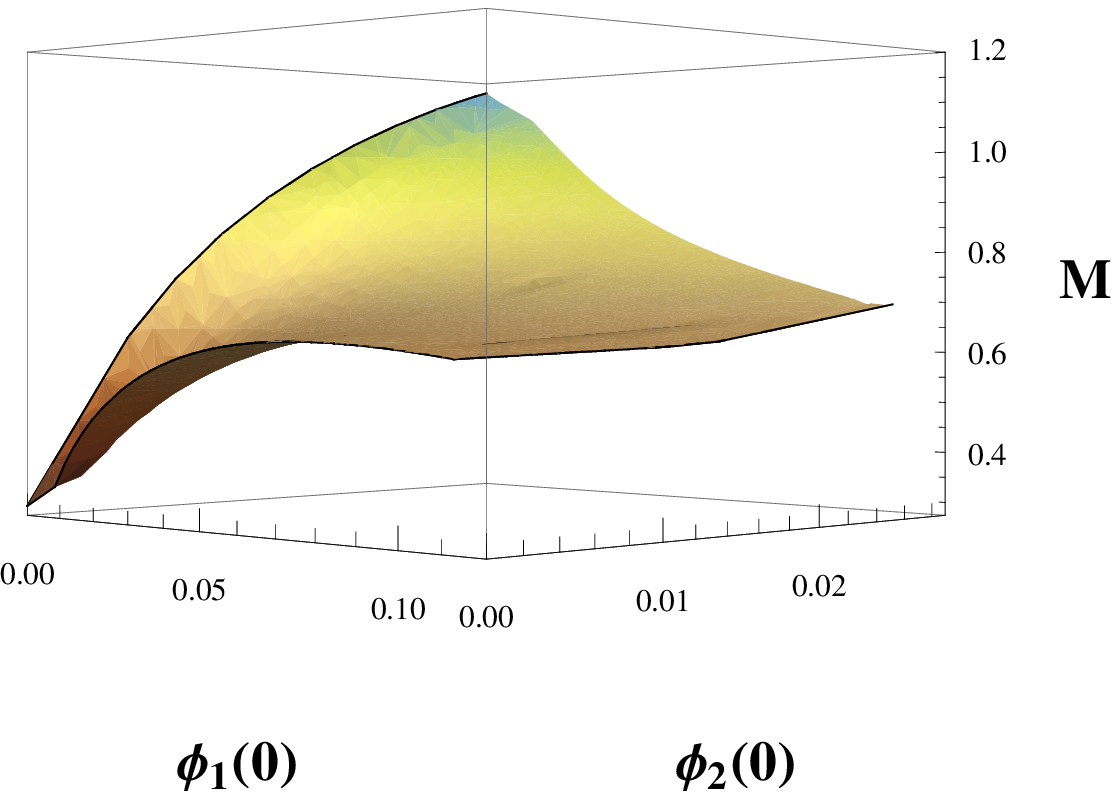}
\includegraphics[width=0.23\textwidth]{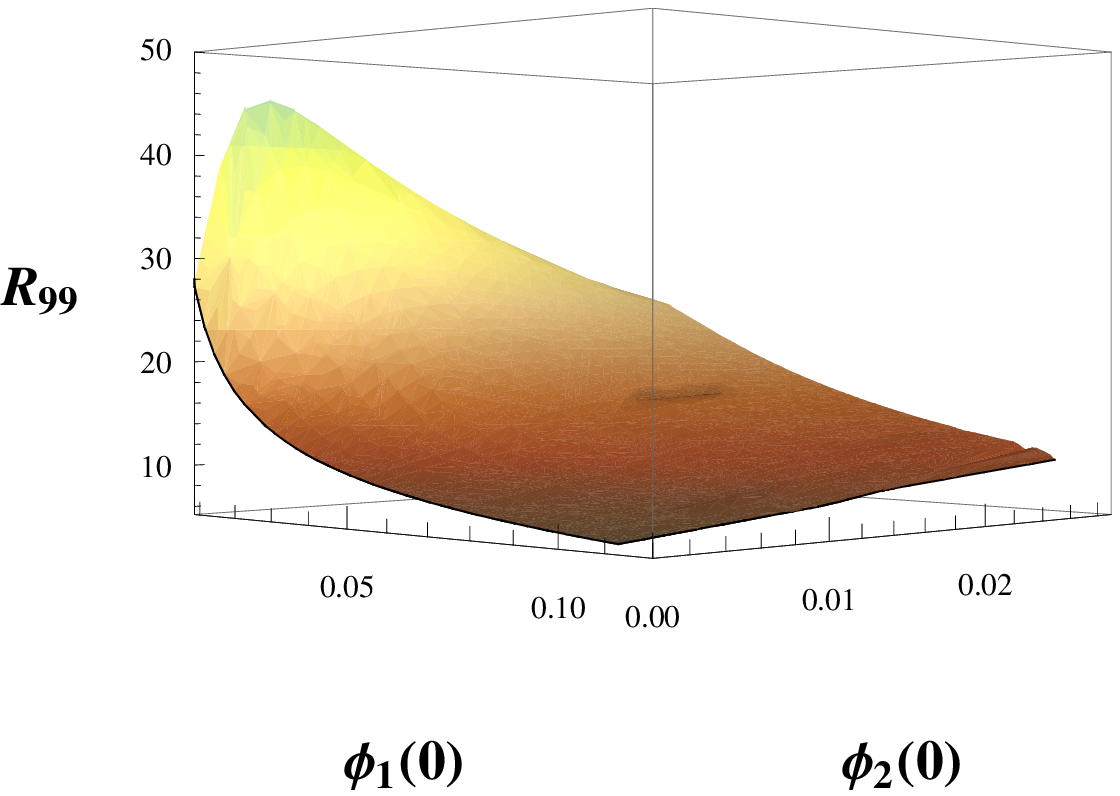}
\caption{Total gravitational mass ($M(\phi_1,\eta)$ ,$R_{99}(\phi_1,\eta)$, 
$M(\phi_1(0),\phi_2(0))$ and $R_{99}(\phi_1(0),\phi_2(0))$) for
initial data of 
MSBS configurations with two states, the ground and the 1st excited state.}\label{Mass-3d-BE}
\end{figure}

In Fig. \ref{Mass-3d-BE} we have plotted the values of these two quantities for all 
the constructed MSBS initial configurations. On the top panel they are shown
as functions of the central value of the scalar 
field in the ground state $\phi_1(0)$ and the Noether fraction $\eta$, 
while in the bottom panel they are plotted as functions of the central value of the 
scalars fields. 

\begin{figure}[h]
\includegraphics[angle=270,width=0.23\textwidth]{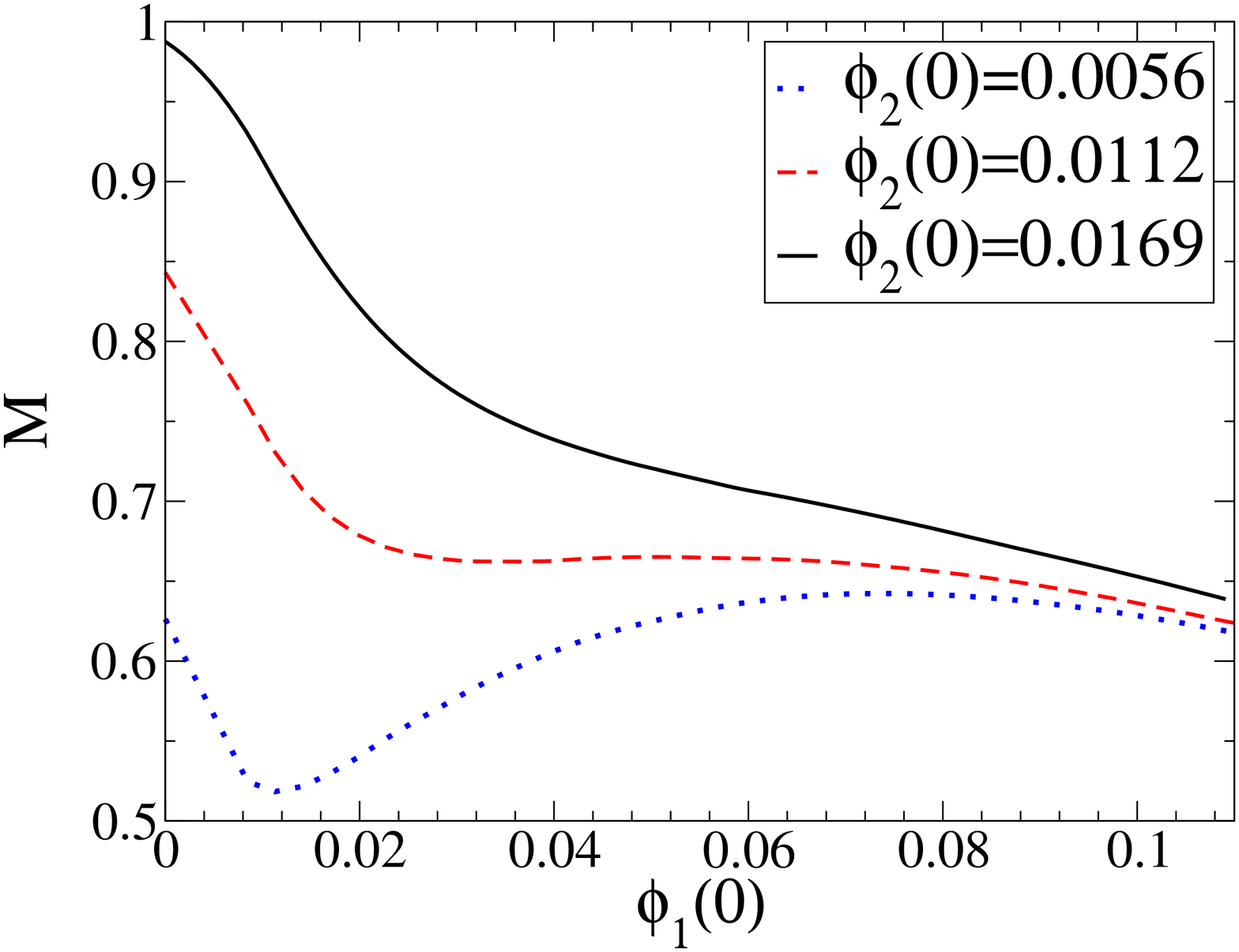}
\includegraphics[angle=270,width=0.23\textwidth]{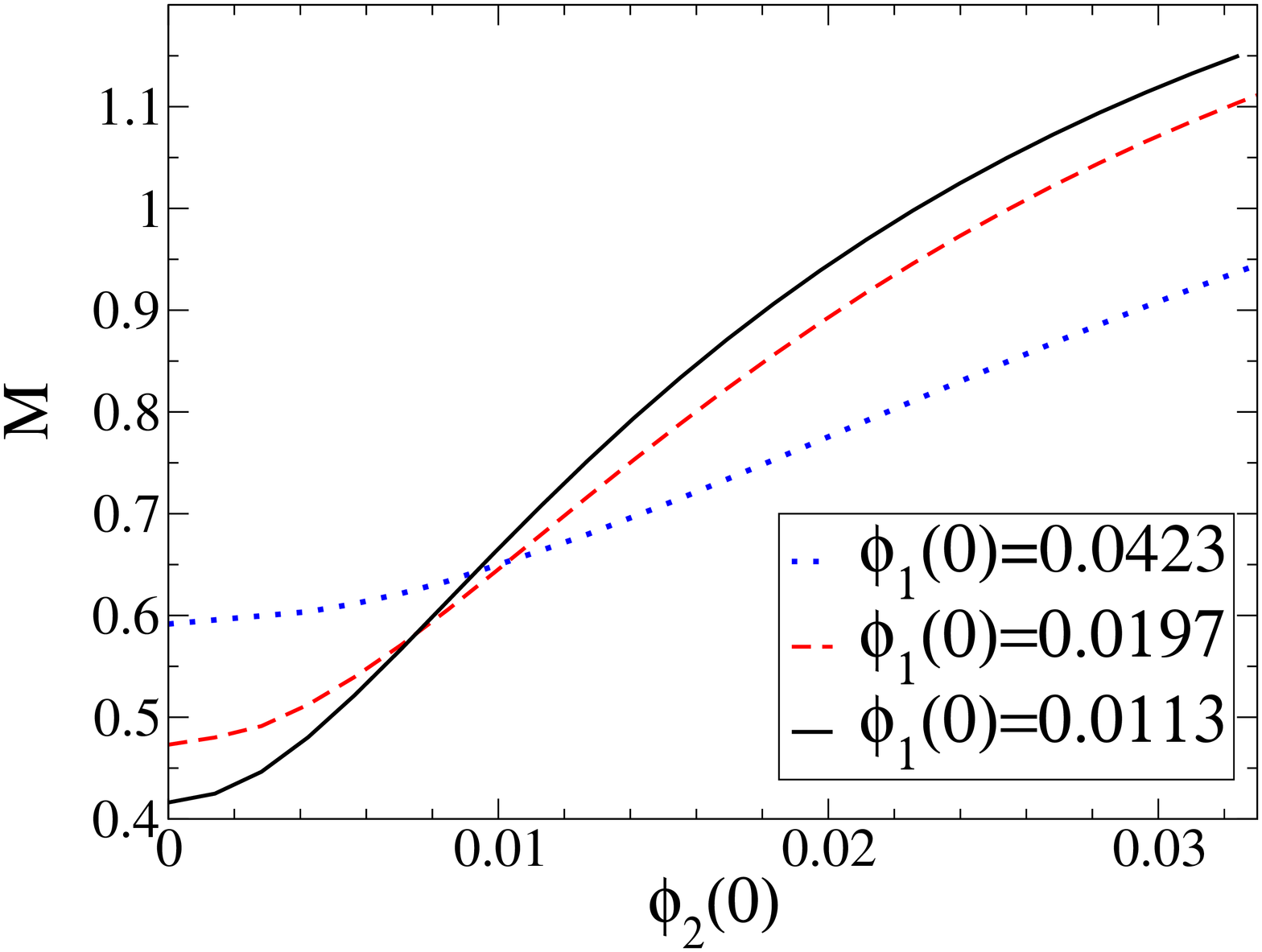}
\includegraphics[angle=270,width=0.23\textwidth]{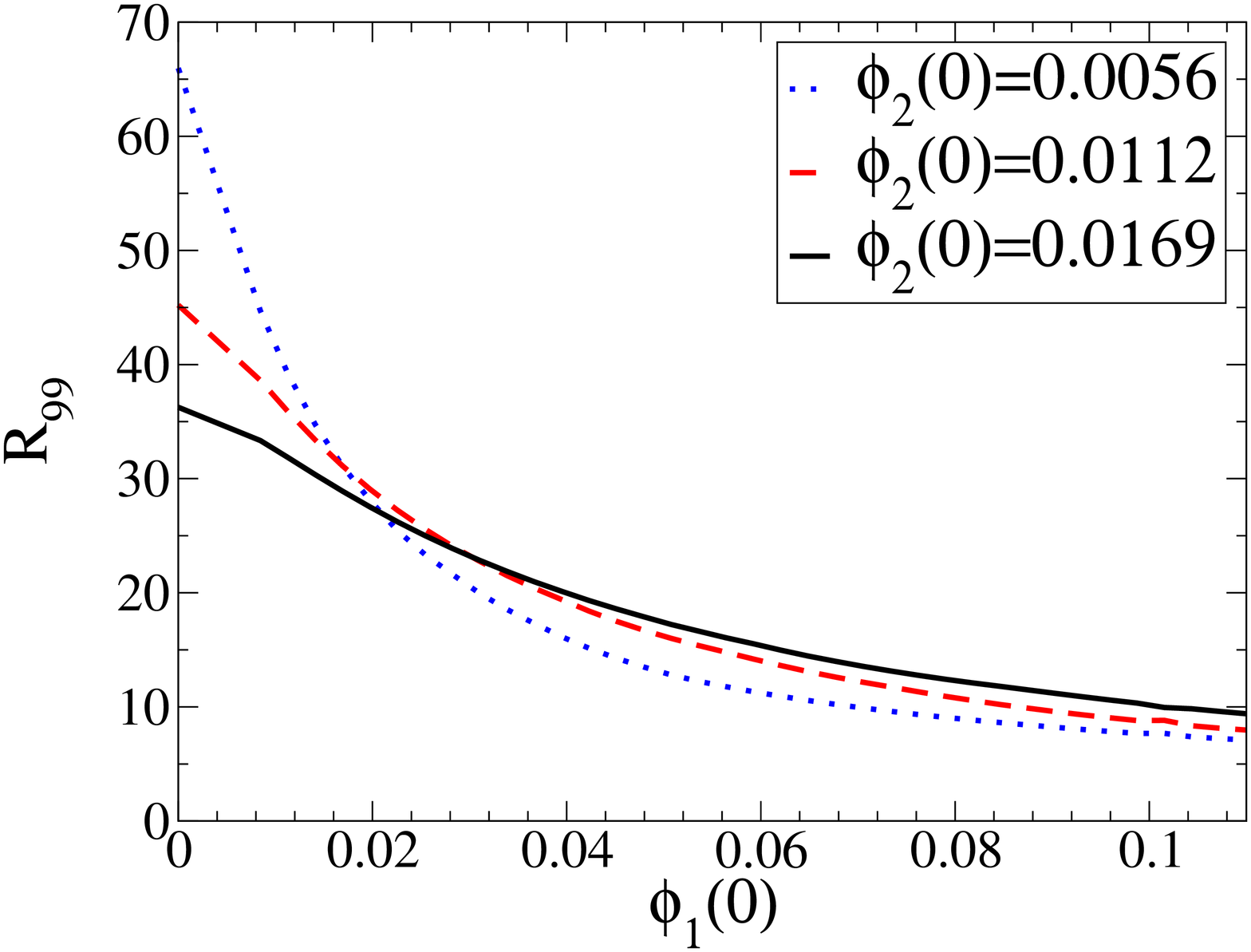}
\includegraphics[angle=270,width=0.23\textwidth]{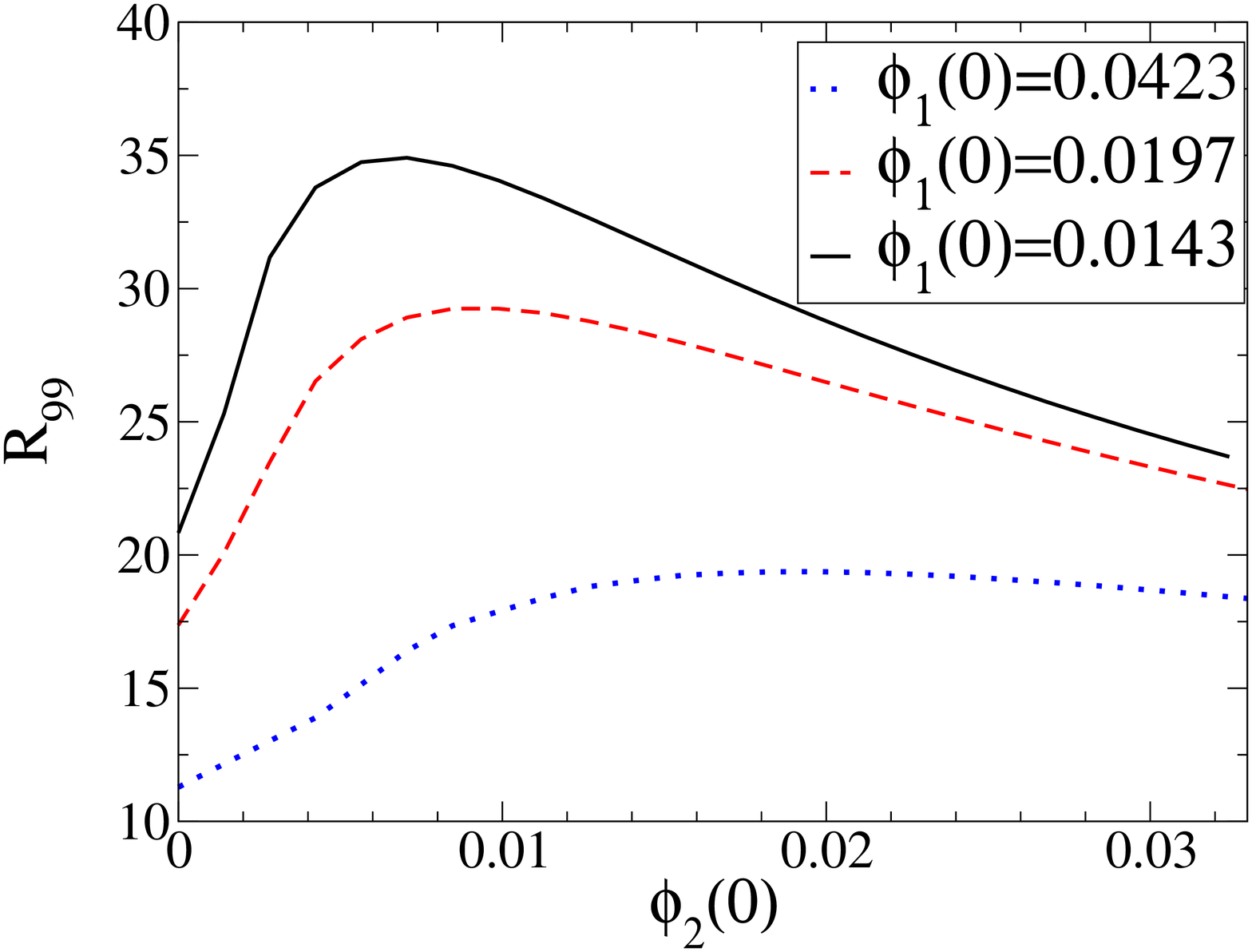}
\caption{Different slices of the Mass (upper panel) and the radius $R_{99}$ 
(lower panel) as a function of the central value of the scalar field.
 On the left there are slices at $\phi_2(0)=constant$ while on the right
there are $\phi_1(0)=constant$.}
\label{MassR-2d-slices}
\end{figure}

These figures already show some differences between single BS and 
MSBS configurations: there are an infinite number of possible equilibrium configurations (i.e. solutions for the 
static EKG system (\ref{ekg1}-\ref{ekg2})) between the two extreme
cases $N^{(1)}=0$ and $N^{(2)}=0$ that correspond precisely to the ground state BS
and first excited state BS respectively.
Fig. \ref{Mass-3d-BE} shows a more complex behavior of the mass $M$ and $R_{99}$ than 
for a single BS, which can be observed easily in Fig. \ref{MassR-2d-slices} where
some slices $\phi_1(0)=constant$ and $\phi_2(0)=constant$ of these surfaces are displayed.
After constructing these configurations, the next problem is stability. It is clear that the known
results of BS stability are not immediately applicable to MSBS.

The stability of the multi-states configurations under finite perturbations 
will be addressed in the next section.

\section{Numerical simulations}
\label{numerical}

In this section, we present a numerical analysis of
the dynamical properties of MSBS, focusing on long-term stability and the final state
of the unstable configurations. To this purpose, we write the 
Einstein-Klein-Gordon system as a set of evolution equations for the scalar
fields and the metric components.
We consider a generic spherically symmetric spacetime with
the line element
\begin{equation}\label{line_element}
ds^{2} = - \alpha^{2}dt^{2} + g_{rr}dr^{2} + r^{2}g_{\theta\theta}d\Omega^{2},
\end{equation}
where $\alpha$ is the lapse function and $\{g_{rr},g_{\theta\theta}\}$ are the 
metric components. Notice the explicit dependence on the factor $r^2$,
such that the component $g_{\theta\theta}$ is regular at the origin. This is
a necessary condition in our implementation in order to deal with the coordinate singularity
at $r=0$.

The evolution equations for the geometry are obtained by substituting the
metric coefficients (\ref{line_element}) into a particular formulation of the Einstein
equations. In this study we have considered the Z3 formulation 
\cite{Bona:2003qn}, which includes the momentum constraint into the evolution
system, by considering an additional vector $Z_i$ as an evolved field. Further details
regarding the Z3 system in spherically symmetry can be found in \cite{Alic:2007ev},
while the regularization of the coordinate singularity $r=0$ is similar
to the one described in \cite{Arbona:1998hu}.

In spherical symmetry, there are independent evolution equations only for the
lapse $\alpha$, the metric components $\{ g_{rr}, g_{\theta \theta}\}$,
the extrinsic curvature $\{ K^r_{r}, K^{\theta}_{\theta}\}$ and
the Z-vector components $\{ Z_r, Z_{\theta} \}$. All these evolution equations
are prescribed by the Einstein equations, except the one corresponding to the lapse,
which is related to the choice of coordinates and can be specified freely.
A common choice, which give rise to a hyperbolic system of equations, is
the harmonic slicing
\begin{equation}\label{eq::lap}
\partial_{t}\alpha = -\alpha^{2} trK,
\end{equation}
where $tr K = K^r_r + 2 ~K^{\theta}_{\theta}$.

On the other hand, the evolution equations for the scalar fields are obtained by
substituting the spherically symmetric metric (\ref{line_element}) in the Klein-Gordon
equations (\ref{KG1}). 

A first order reduction in space can be performed by introducing as
independent quantities the
spatial derivatives of the metric and new fields related to the time and spatial
derivatives of the scalar field, namely
\begin{eqnarray}
   A_r &\equiv& \frac{1}{\alpha}\partial_r \alpha ~,~
   {D_{rr}}^{r} \equiv \frac{g^{rr}}{2} \partial_r g_{rr} ~,~
   {D_{r\theta}}^{\theta} \equiv \frac{g^{\theta\theta}}{2}
                        \partial_r g_{\theta\theta} , 
\nonumber \\
  \phi^{(n)}_{r} &\equiv& \partial_r \phi^{(n)} ~,~
  \phi^{(n)}_{t} \equiv \frac{\sqrt{g_{rr}}}{\alpha} \partial_t \phi^{(n)}.
\end{eqnarray}

In this way, we obtain a fully first order system of evolution equations
for the geometry and the scalar fields, with the following set of evolution variables
\begin{equation}
 \{ \alpha, g_{rr}, g_{\theta \theta},
K^r_{r}, K^{\theta}_{\theta},
A_r, {D_{rr}}^{r}, {D_{r\theta}}^{\theta}, Z_{r}, \phi^{(n)}, \phi_r^{(n)},
 \phi_t^{(n)}\}. \nonumber
\end{equation}
This first order system can be written in balance law form
\begin{equation}
     \partial_t {\bf U} + \partial_k ~ ^k F({\bf U}) = S({\bf U}),
\end{equation}
which allows the use of advanced numerical methods based on Finite Volume algorithms.
Details about the exact form of the evolution equations
can be found in Appendix \ref{the_Z3_system}.

We have implemented the equations using the
Method of Lines, in order to separate the time and the spatial
discretization. The time integration is performed with a third order
Strong Stability Preserving Runge Kutta method \cite{ShuOsher88}.
The spatial discretization is based on a standard fourth order centered finite difference scheme, plus third order accurate dissipation \cite{Alic:2007ev}.


\subsection{Stability of the MSBS}
\label{stability}

The stability of MSBS configurations is a basic requirement for considering
them suitable models of galaxy halos. For a single boson star, the stability has been previously
studied both analytically and numerically, showing that it is stable if
$\phi(0) \le \phi_{max}(0)$. In this section, we analyze the stability of MSBS
in the range $\phi_1(0) \le \phi_{max}(0)$, which are stable for $\eta=0$,
avoiding this way the too-massive unstable MSBS. The stability will be studied
following only a numerical approach, by perturbing the MSBS and studying the evolution of this
perturbation. We will restrict ourselves to study stability against spherically symmetric
perturbations by using the equations described in the previous subsection.
Notice that this is only a necessary condition for the most general case,
since asymmetric perturbations may still be unstable.

In order to study numerically the stability of the MSBS configurations, we perform
the following steps:
\begin{itemize}
  \item Construct different initial data sets for MSBS with a given $\phi_1(0)$,
        by varying the Noether fraction $\eta$. 
  \item Add a real scalar field far outside the radius $R_{99}$ of the MSBS,
        which will be coupled to the MSBS only through gravity.
        The energy density corresponding to the scalar field is only $0.01\%$
        of the total energy density of the MSBS,
        so it will act just as a small perturbation with negligible errors 
        in the form of constraint violations.
  \item Perform evolutions of the Einstein-Klein-Gordon system and study
        the behavior of the MSBS. The scalar field perturbation will fall into the
        MSBS and later disperse to infinity. The gravitational interaction during
        that time is expected to excite the unstable modes, if any. In this
        way, the modes are excited sooner than only by numerical errors.
  \item Bracket the MSBS which lead to significant exponential growing modes.
        For the stable MSBS configurations, the perturbations will only oscillate
        without growing. We expect that the MSBS with low $\eta$ will be stable,
        since the major contribution to the complete configuration comes from the stable
        ground state, while those with high $\eta$ will correspond to unstable MSBS.
  \item Fit the growth rate of the unstable MSBS for each set of stars with the
        same $\phi_1(0)$ by varying $\eta$. Extrapolate to find the maximum allowed
        Noether fraction $\eta_{max}$ which separates the stable and unstable states.
        This procedure allows us to obtain reliable estimations, without evolving every configuration in order
        to obtain $\eta_{max}$. Moreover, it might be difficult to distinguish
        stable from unstable configurations when they are close to $\eta_{max}$,
        since the exponential growth is very low in that region.
\end{itemize}

\begin{figure}[h]
\includegraphics[angle=-90,width=0.4\textwidth]{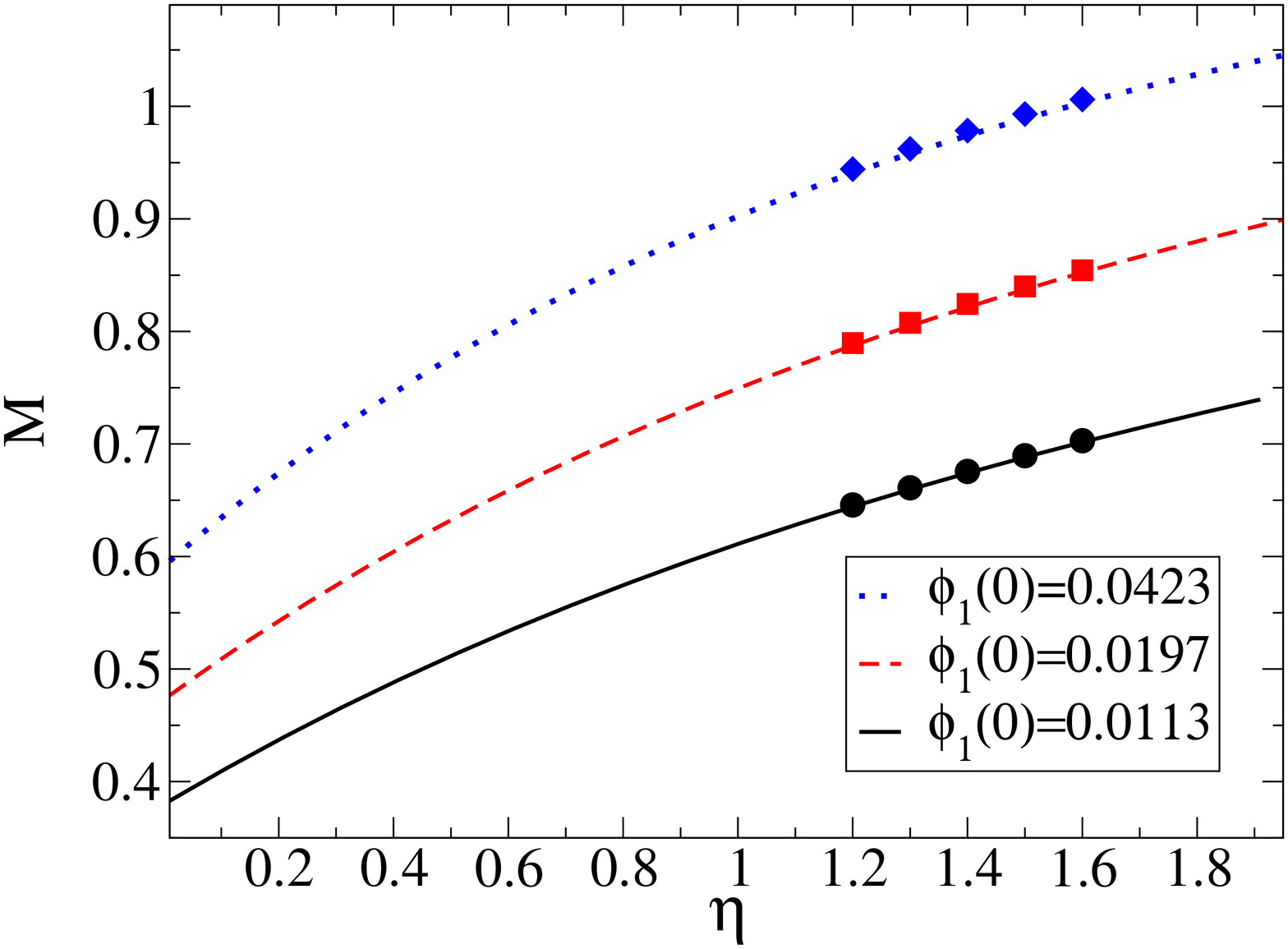}
\includegraphics[angle=-90,width=0.4\textwidth]{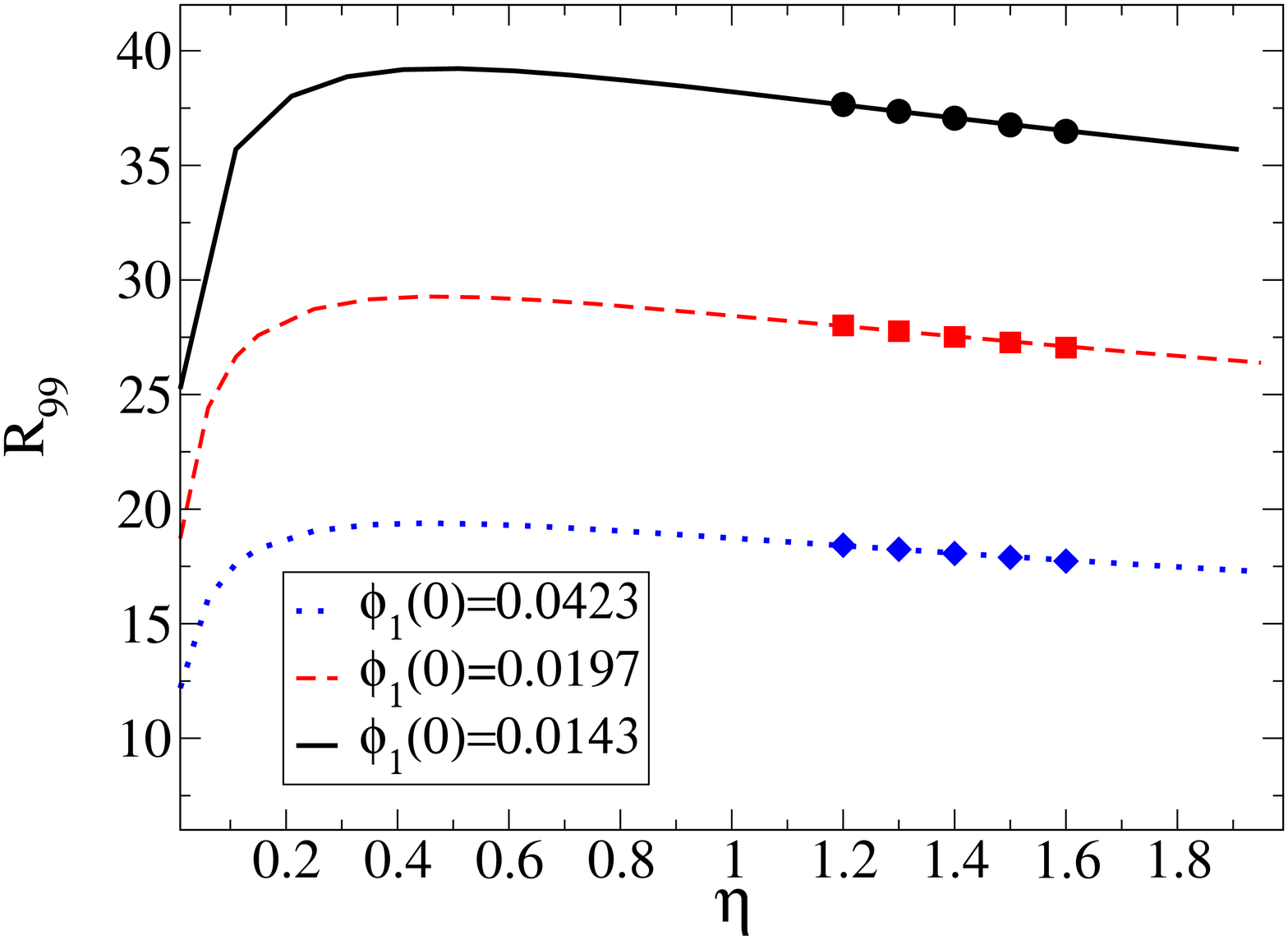}
\caption{The Mass (upper panel) and the Radius $R_{99}$ (lower panel) are
  presented as a function of $\eta$, for three
different central values of the ground state field $\phi_1(0)$. The points
are the configurations used for the fitting of $\eta_{max}$.}
\label{MassR-2d}
\end{figure}

We are going to restrict the numerical stability analysis to only three different
values of $\phi_1(0) = \left\{0.0143, 0.0197, 0.0423 \right\}$. Fig. \ref{MassR-2d} displays
the total mass and the radius $R_{99}$ for these configurations, as a function
of $\eta$. In the simulations with $\eta \le 1.2$, we did not detect any unstable
exponentially growing mode, or they were difficult to measure for some families
of solutions. The results indicate as the upper bound $\eta_{max} <
1.2$.

\begin{figure}[hbtp]
\includegraphics[angle=0,width=0.42\textwidth]{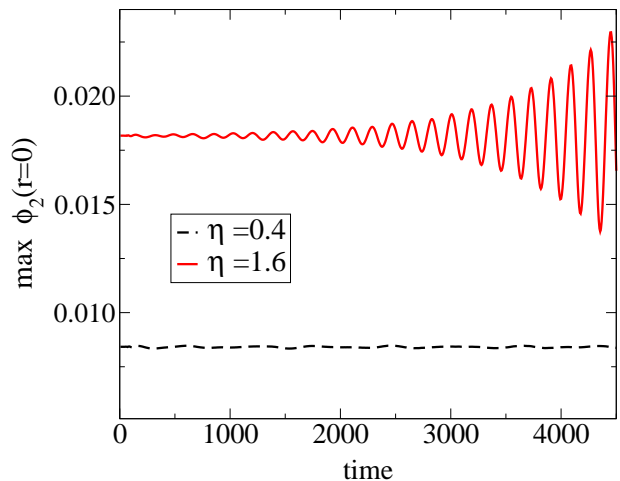}
\caption{Maximum of the central value of the scalar field in the excited
state $\phi_2$, for two different values of $\eta$ with $\phi_1(0)=0.0197$.
The MSBS with $\eta =0.4$ is in the stable branch and the induced perturbations
do not grow. The MSBS with $\eta=1.6$ is clearly unstable and the perturbations
exhibit an exponential growth.}
\label{maxphi2}
\end{figure}

In fig. \ref{maxphi2}, we show the typical behavior for one of these simulations,
corresponding to stable and unstable MSBS. The perturbation has an exponentially
growing behavior only for the unstable MSBS. The maximum of the scalar field
$\phi_2(r=0)$ can be fitted with a function of the type
\begin{equation}\label{fitting}
     A ~ \exp( \sigma t) ~ cos(\omega t + \varphi),
\end{equation}
which allows us to compute the exponential growth rate $\sigma$.

\begin{figure}[h]
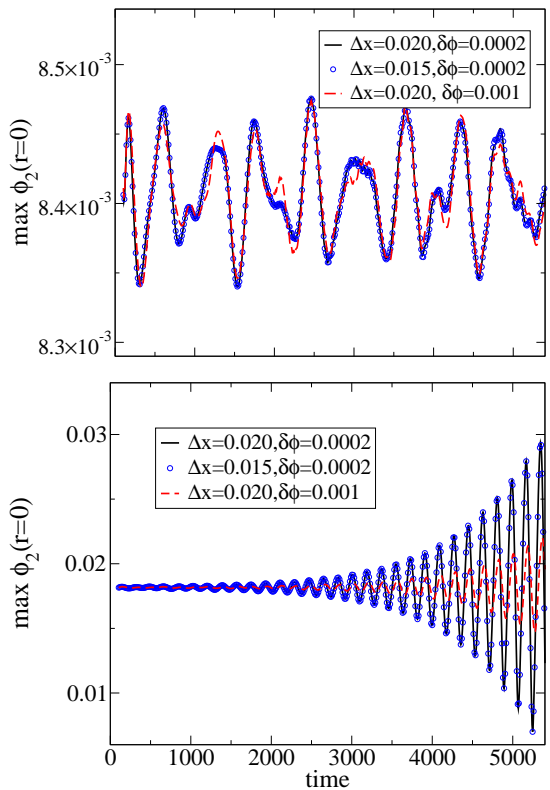

\includegraphics[angle=0,width=0.4\textwidth]{phi2_04.eps}
\includegraphics[angle=0,width=0.4\textwidth]{phi2_16.eps}
\caption{The same as in fig.\ref{maxphi2}, for different resolutions and perturbation
amplitudes. Despite a delay in the excitement of the exponential mode of the unstable
case, the results are robust with respect to changes in resolution and perturbation.}
\label{maxphi2b}
\end{figure}

\begin{figure}[hbtp]
\includegraphics[angle=0,width=0.4\textwidth]{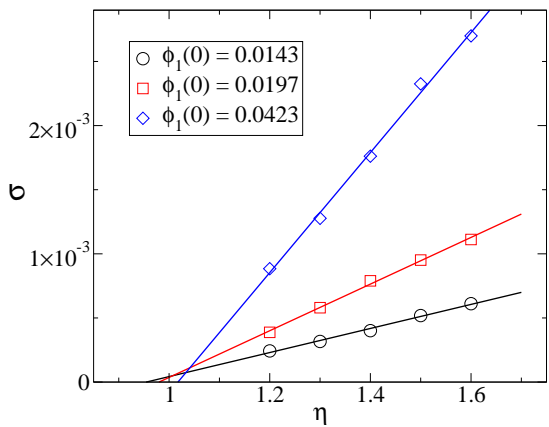}
\caption{Fitting of the exponential growth for three families of MSBS configurations. The
   extrapolated value corresponds to the maximum allowed stable fraction.}
\label{fit}
\end{figure}

We performed fits for the unstable MSBS perturbations with
$\eta \ge 1.2$, marked with filled geometrical shapes in fig. \ref{MassR-2d}. The results
for $\sigma$ are represented in fig. \ref{fit}, with the extrapolation to the
$\eta_{max}$, which in principle could be a function of $\phi_1(0)$. The three
different families of solutions point to $\eta_{max} \approx 1$.

In order to show the robustness of these results we have repeated the 
simulations with more resolution and with a different amplitude of the perturbation.
The results are almost identical, as shown in fig.\ref{maxphi2b}. The only
significant difference is the unstable case with larger perturbation amplitude. 
The perturbation seem to interact non-linearly with the star, and the unstable
exponentially growing mode is excited later. In spite of this delay, the growth rate is identical 
to the other cases with smaller perturbation and higher resolution.


\subsection{Fate of the unstable states}
\label{unstable states}

Another question which arises from the previous stability analysis, refers 
to the final fate of the unstable MSBS with $\eta > \eta_{max}$.
We address this issue by performing long evolutions of
unstable MSBS configurations, until they reach a stationary
state. In order to accelerate the growth of the unstable modes,
we perturbed the MSBS with a massless scalar field located far
from $R_{99}$. As explained in the previous section, this scalar
field interacts gravitationally with the MSBS, perturbing it slightly and
exciting the unstable modes. These modes grow exponentially,
starting with a small amplitude, result that can be obtained also 
from a linear perturbation analysis. When the amplitude of these perturbations
is larger, the nonlinear effects become important and the evolution
can only be followed numerically in order to discern the final state
of the MSBS. 

\begin{figure}[h]
\includegraphics[angle=0,width=0.40\textwidth]{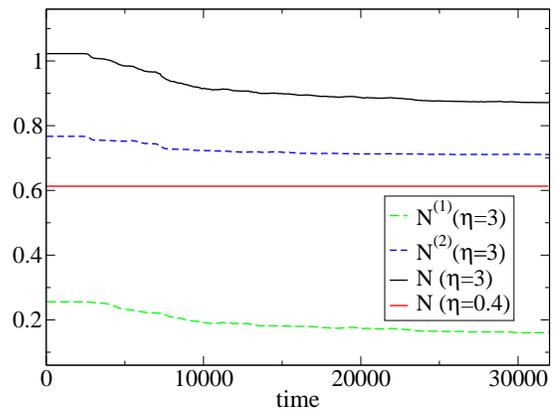}
  \caption{The number of particles in the ground state and in the excited
state for the fraction $\eta=3$, together with the total number of particles
for the fraction $\eta=3$ and $\eta=0.4$. There is a significant loss of the 
number of particles in the unstable configuration, while
the stable only looses $0.004\%$ due to numerical dissipation.}
\label{Ns}
\end{figure}

\begin{figure}[h]
\includegraphics[angle=0,width=0.40\textwidth]{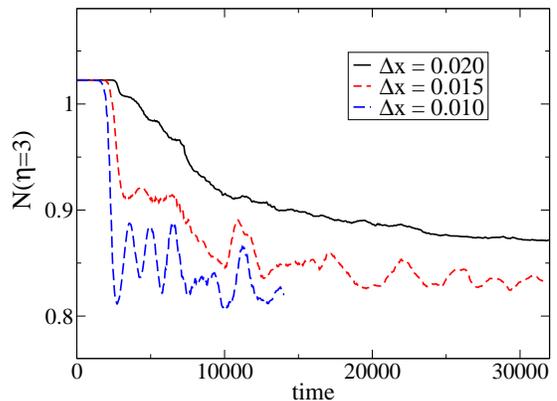}
  \caption{The total number of particles for the fraction $\eta=3$ computed with three different resolutions. The loss in the 
number of particles converges to a value around $18\%$.}
\label{NT3}
\end{figure}

\begin{figure}[h]
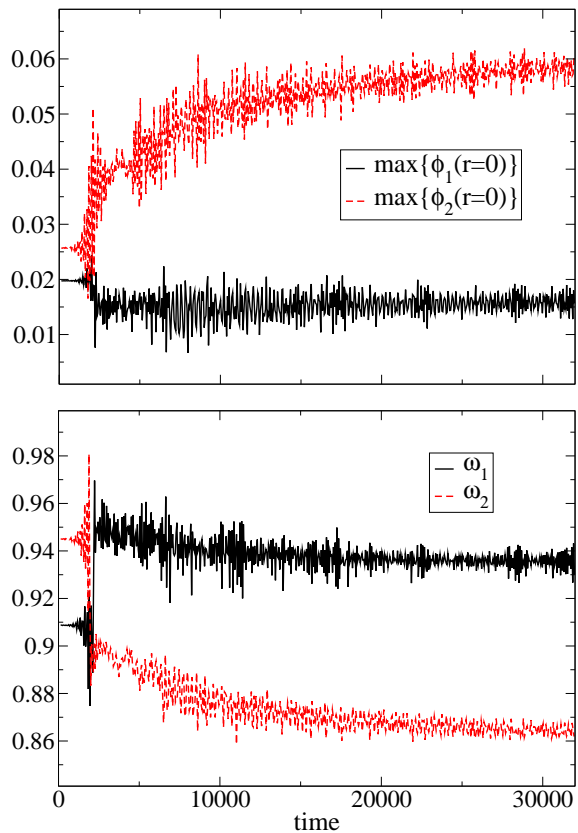

\includegraphics[angle=0,width=0.42\textwidth]{phi3.eps}
\includegraphics[angle=0,width=0.42\textwidth]{freq3.eps}
  \caption{The maximum of the central value of the scalar fields in the different states
for the fraction $\eta=3$ on the top, and the frequencies of those modes on the bottom.
There is a flip-flop of the scalar field and the frequencies at $t=5000$.}
\label{phis}
\end{figure}

Fig. \ref{Ns} displays the number of particles in different states and the total number of particles
for two MSBS with the same $\phi_1(0)=0.0197$ with a fraction given by $\eta=3$ and $\eta=0.4$.
The total Noether charge remains almost constant
in the stable case $\eta=0.4$, showing the accuracy of the numerical code within
a $0.004\%$ error in this quantity. The unstable case $\eta=3$ exhibits scalar
field radiation during the
evolution, producing a decrease of around $18\%$
in the total Noether charge, as it can be seen in the convergence test presented in fig.~\ref{NT3}. This radiation translates into a small change
in the amplitude of the scalar fields. Taking a closer look at the maximum
value of the scalar fields in the center and the frequencies, displayed in figure
\ref{phis}, one can notice a change in the position of the node; the excited state has decayed to a ground one, while the ground one has jumped to the first excited state. With this ``flip-flop'' of the
scalar fields, the final $\eta$ is in the stable domain. A similar behavior
is observed for all the unstable MSBS configurations included in the study, which indicates
that this could be a common feature of their evolution. The time
required by an unstable MSBS to settle down into
a stable configuration, increases as the fraction gets closer to $\eta_{max}$.

\begin{figure}[h]
\includegraphics[angle=0,width=0.42\textwidth]{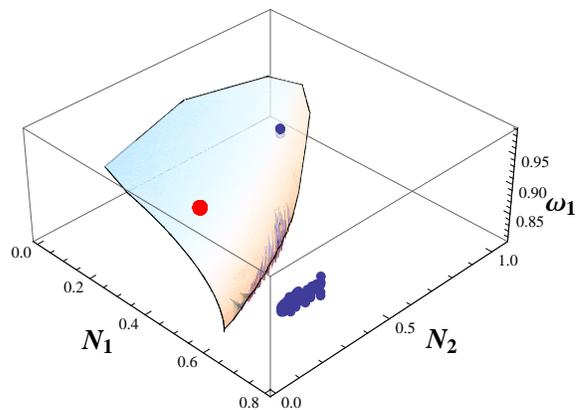}
  \caption{``Stable'' and ``unstable'' states}
\label{distintos}
\end{figure}

The previous statements can be seen graphically in Fig. \ref{distintos}. 
The figure shows the collection of initial configurations, each one
characterized by the number of particles in the ground state $N_1$, and the
number of particles in the excited state $N_2$. Every pair $(N_1,N_2)$ 
corresponds to a configuration with a pair of eigen-values  $(\omega_1,\omega_2)$,
although only $\omega_1$ is shown in the figure. On the top 
of these collection of initial configurations, we have plotted 
the time evolution of the previous simulations.
The red dot corresponds to the configuration labeled with fraction $\eta=0.4$
and the blue dots represent the time evolution for the configuration labeled 
with fraction $\eta=3.0$. Applying the same perturbation for the two different states, we can discern two different behaviors: the fraction $\eta=0.4$ is
stable and remains static, while the configuration with $\eta=3.0$ is 
unstable and evolves far from the original configuration (that
lies on the sheet of static configurations) to a ``forbidden'' region.
The jump is due to the flip of ground state $\to$ excited state and
viceversa. Then, it starts to lose scalar field (i.e., evolves in $(N_1,N_2)$)
and slowly approaches an equilibrium configuration on the sheet of
equilibrium configurations. 

The final state of the MSBS with $\eta=3$ can be inferred from the central
value of the amplitude of the scalar fields (top panel Fig. \ref{phis}).
It is important to remind that now, due to the switch of the frequencies,
the new ground state corresponds to $\phi_2(0)$ while the new excited state corresponds to $\phi_1(0)$. The amplitudes oscillate around a central value,
namely
\begin{eqnarray}
\phi_1(0)^{new}&=&\phi_2(0,t_{final})=0.058\pm0.004 \nonumber\\
\phi_2(0)^{new}&=&\phi_1(0,t_{final})=0.016\pm0.004\,.
\end{eqnarray}
Using these values as initial conditions for the new equilibrium configuration
we can compute the eigenvalues and the number of particles shown in Table \ref{comparacion}.
In the same table are displayed the values obtained from the evolution of the
MSBS configuration with $\eta=3$. There is a remarkable agreement
in the eigenvalues, but they still differ in the number of particles. 
This means that the system will still loose particles at a slow rate during the evolution, as corroborated by Fig. \ref{distintos}. 
The slow loss of particles has been observed in finite perturbed system of excited 
states and even tough the final state has been inferred in the same way as we have done here
\cite{1990PhRvD..42..384S,1998PhRvD..58j4004B}. 

\begin{table}
\begin{tabular}{|c|c|c|c|c|}
\hline
& $\omega_1$ & $\omega_2$& $N^{(1)}$ & $N^{(2)}$\\
\hline
Equilibrium & 0.87 &0.94 & 0.604 & 0.126  \\
Evolved $\Delta$ x = 0.020 &$0.864 \pm 0.004$ & $0.936 \pm 0.006$ &0.71 &0.16 \\
Evolved $\Delta$ x = 0.015 &$0.862 \pm 0.003$ & $0.934 \pm 0.010$ &0.706 &0.126 \\
Evolved $\Delta$ x = 0.010 &$0.872 \pm 0.010$ & $0.940 \pm 0.005$ &0.696 &0.124 \\
\hline
\end{tabular}
\caption{Expected eigen-values and number of particles for the "new" equilibrium configurations.
The values obtained from the late time evolution of the configuration with $\eta=3$ is shown
for comparison.}
\label{comparacion}
\end{table}

\section{Dark matter halos}
\label{DMhalos}

In previous sections we have constructed MSBS with two states, a ground and a first excited state, and shown their stability. In this section, we come back to our initial motivation
and we will illustrate how MSBS can lead to RCs which are in better agreement the RCs of galaxies in the context of 
SFDM.
A detailed analysis of fitting the RC of galaxies including baryonic matter and 
experimental data is out of the scope of the present work. 
Instead, we will just present a comparison between the behavior of a test particle
immersed in the gravitational potential
produced by a MSBS and a single BS, showing that there is an improvement in the
sense that the MSBS has a flatter profile far from the center.Neglecting the baryonic contribution is a 
reasonable assumption in some galaxies such as the Low Surface Brightness Galaxies.

For the static spherically symmetric metric considered here (\ref{line_element}), the circular
orbit geodesic obeys \cite{1973grav.book.....M},
\begin{equation}
v_\varphi^2=r \alpha\, \partial_r \alpha \,\,.
\end{equation}

\begin{figure}[hbtp]
\includegraphics[angle=270,width=0.4\textwidth]{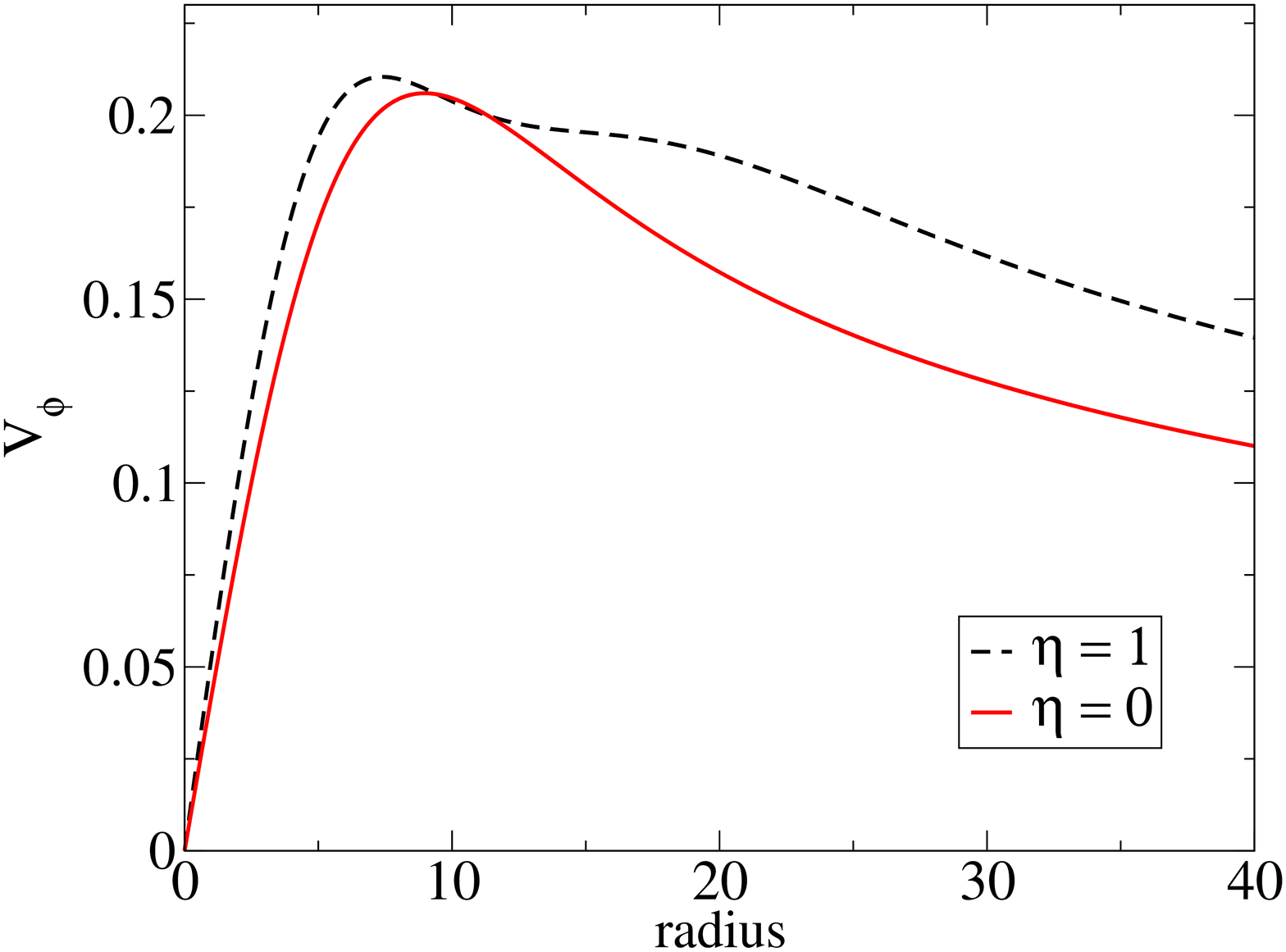}
\includegraphics[angle=270,width=0.4\textwidth]{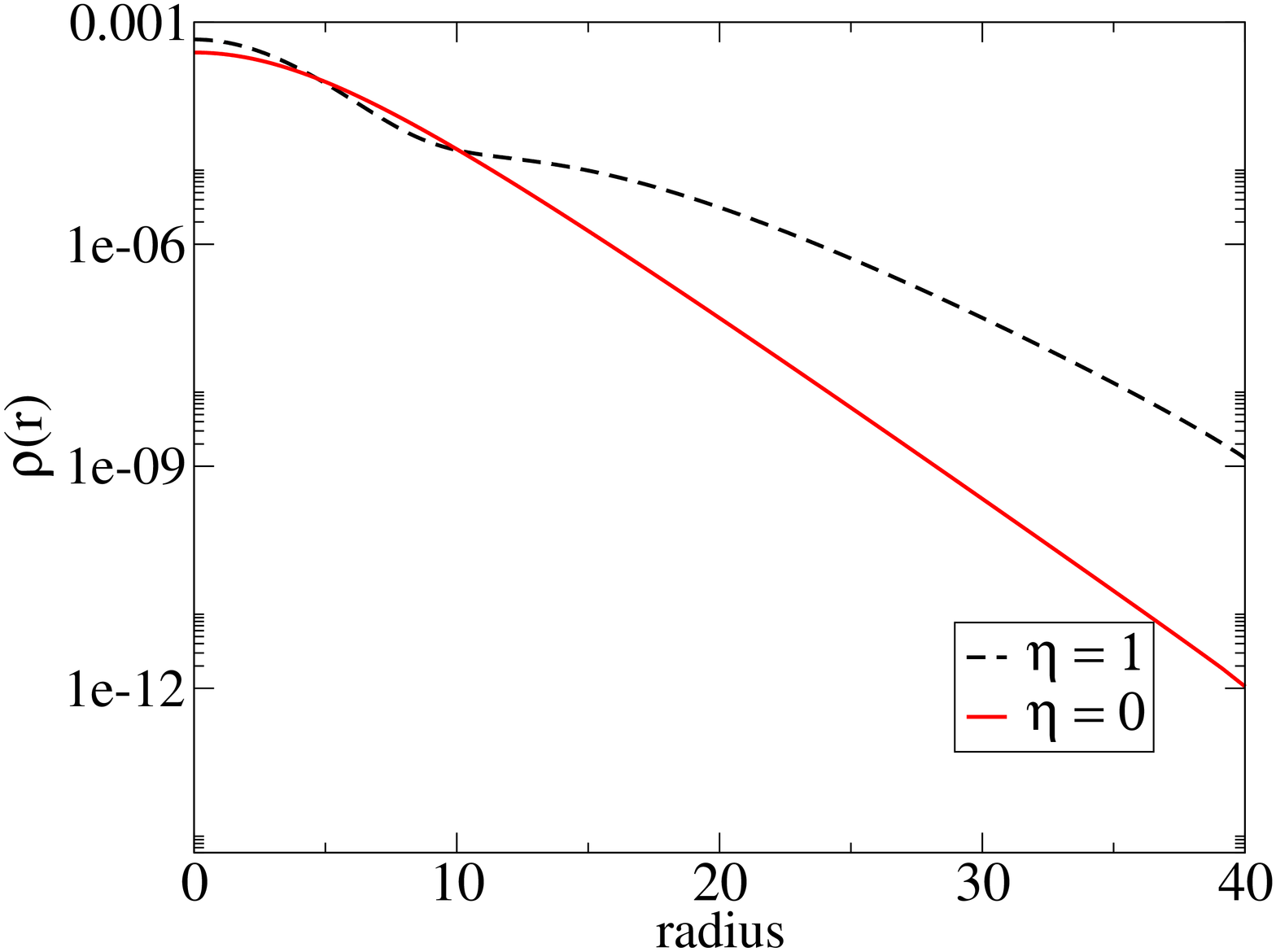}
\caption{Density profiles and rotational curves for a MSBS with 
$\phi_1(0)=0.0197$ and fraction $\eta=1$, and the standard Boson-star 
made of a single state scalar field with the same central value 
$\phi_1(0)=0.0197$ ($\eta=0$). Improvement in the keplerian tale is observed 
for large values of $r$.}\label{rotationalcurve}
\end{figure}

As an illustrative example, the top panel of Fig. \ref{rotationalcurve} presents 
a comparison between two rotational
curves obtained for a single state boson star with $\phi_1(0)=0.0197$ (ie, a MSBS with $\eta=0$)
and for a MSBS with the same amplitude $\phi_1(0)=0.0197$
of the ground state scalar field, and the same number of particles in
the first excited state (ie, $\eta=1$). The region with a flat plateau is larger,
suggesting that for higher excited states (or MSBS with several
higher states) the region with constant rotational velocities could be extended to
larger radii.

In order to understand this behavior, it will be helpful to see the mass density profile
defined as
\begin{equation}\label{mass-density}
\rho(r)=\frac{1}{r^2}\frac{dM(r)}{dr}\,,
\end{equation}
and it is shown in the bottom panel of Fig. \ref{rotationalcurve} for the same configurations
mentioned above.
We can see that for the single BS ($\eta=0$) the density decays exponentially as $r \to \infty$, making it difficult to fit the flat rotational curve
profiles present in most galaxies.
However, the MSBS configurations with large Noether fractions have a radius which is significantly
larger than the one corresponding to the single ground state, with an exponential decay
only in the tail of the excited state. 

Another issue related with boson stars as dark matter models, was the lack of degrees of
freedom to match the different sizes and masses of the observed galaxies. For a
single boson star without self-interaction, the only free parameters are the mass
of the boson particle $m$ and the central value of the scalar field $\phi(r=0)$, which
determines the compactness of the object (ie, ratio of total mass over radius)
in adimensional units. There have been several attempts to fit these parameters 
\cite{1994PhRvD..50.3650S,Ji:1994xh,Guzman:2006yc,Arbey:2003sj,Matos:2007zza} 
with different levels of success. By allowing more general MSBS, there are
extra free parameters coming from the different fractions between the ground
and excited states. These parameters change not only the total mass, but
also the compactness of the final object. The extra degrees of freedom may allow a better fit 
of the models to different galaxies.
\section{Conclusions}
\label{conclusions}

\begin{figure}[hbtp]
\includegraphics[angle=270,width=0.49\textwidth]{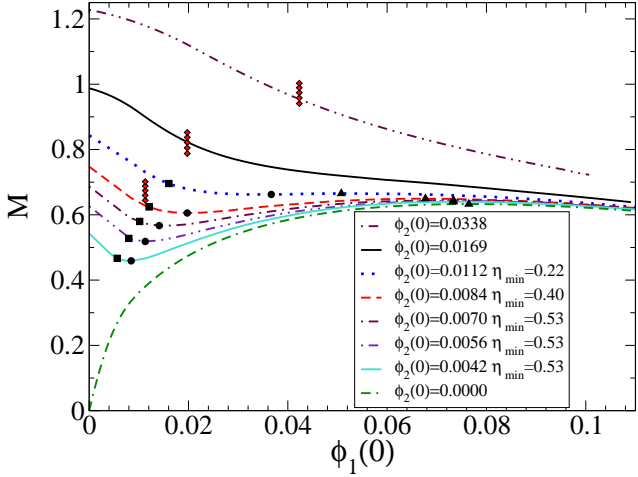}
\caption{Mass as a function of $\phi_1(0)$ for different values of $\phi_2(0)=constant$. The triangles mark the maximum allowed masses (ie, the maximum of the curves) while the circles correspond to the minimums of the curves. The maximum fraction $\eta_{max}$, displayed
with squares, is found numerically on the left of the minimum of the curves.}
\label{masa-phi1-phi2fix}
\end{figure}

We have constructed generalized boson star configurations, where two coexisting states 
of the scalar field are present. Our initial data construction is based on two main quantities that
describe them: the gravitational mass $M$ and the radius $R_{99}$. 
We have shown that these boson stars are stable under small radial perturbations, for a certain range of the fraction ($\eta<1$) between the Noether charges. 
These results may sound counter-intuitive, given the known fact that
single BS in excited states are unstable under finite perturbations. 
Nevertheless, the addition of an extra scalar field allows for an infinite number of new equilibrium configurations. The known plot of $M$ vs. $\phi_1(0)$ for ground state or excited state single BSs is now extended, getting a different curve for each fixed value of $\phi_2(0)=constant$, as it is shown in the figure~\ref{masa-phi1-phi2fix}. The single BS in the ground state corresponds to the case $\phi_2(0)=0$, which only has an extreme at the maximum allowed mass. The configurations on the right of that point (marked with a triangle) are
unstable. For low values of $\phi_2(0)$ these curves contain now two extremes.
In addition to the maximum allowed mass, there is a minimum close to
the fraction $\eta \approx 0.5$. Although the presence of a new extreme in the curves could
suggest a change in the stability around the fraction $\eta \approx 0.5$, our
numerical stability analysis presented in the section \ref{stability} indicates
that the dynamical stability regime is extended beyond this point
up to $\eta_{max} \approx 1$. Furthermore, we have found also stable and unstable 
configurations in regions with no extremes in the figure~\ref{masa-phi1-phi2fix}. This
supports the idea that simple stability arguments from the single boson star case
are not valid in this case and that a more detailed future study is necessary in order to confirm and understand completely this problem.

The unstable configurations evolve and settle down into stable configurations. 
MSBS allow to obtain a flat region in the velocity rotational curves,
as shown by the examples in this paper. We considered cases with only two
different states of the scalar field. 
As future work, we are planning to construct
MSBS where several states are coexisting. These models allow more degrees
of freedom, and could be used  
to fit accurately the rotational curves within the observational data.

\acknowledgments
We acknowledge Shin Yoshida for invaluable discussion and Steve Liebling, Luis Lehner,
Carles Bona, L. Urena, F.S. Guzman and Bruno Giacomazzo for useful comments. AB, JB 
and DA thank L. Rezzolla for his support and hospitality at the AEI.

This work was supported in part by the CONACYT and CONACYT-SNI, the Spanish Ministry of
Science and Education under the FPI fellowship BES-2005-10633 and the
DFG grant SFB/Transregio 7.

\appendix




\section{Equivalence of real quantized scalar field and multi boson stars}
\label{quantized}

The many boson-system is described by a second quantized free scalar field
\begin{equation}\label{quantizedphi}
\hat \Phi=\sum_{nlm}\hat b_{nlm}\Phi_{nlm}(t,{\bf x})+
\hat b_{nlm}^\dagger \Phi_{nlm}^*(t,{\bf x}),
\end{equation}
 with an energy-momentum tensor operator given by
\begin{equation}\label{quantizedt}
\hat T_{ab}=\partial_{a}\hat \Phi \partial_{b}\hat\Phi-\frac{1}{2}g_{a b}(
g^{c d}\partial_{c}\hat\Phi \partial_{d}\hat\Phi+\mu^2|\hat\Phi|^2),
\end{equation}
where $\mu$ is the scalar field mass and the convention $\hbar=c=1$ has been adopted.
The gravitational field is treated as a classical field, so the source in the 
r.h.s. of the Einstein equations has the expectation value of 
(\ref{quantizedt})
over a state of the system of many particles $|Q\rangle$,
namely
\begin{equation}\label{einsteineq}
G_{ab}=8\pi  \langle Q| \hat T_{ab} |Q\rangle.
\end{equation}

The operators in the quantized field expansion of eq. (\ref{quantizedphi})
can be interpreted as creation  
$\hat b_{nlm}$ and annihilation $\hat b_{nlm}^\dagger$ quantum operators. 
These operators satisfy the following commutation relations:
\begin{equation}\label{conmutators}
[ \hat b_{nlm},\hat b^\dagger_{n'l'm'}]=\delta_{nn'}\delta_{ll'}\delta_{mm'} \,, 
\end{equation}
\begin{equation}
[ \hat b_{nlm},\hat b_{nlm}]=[\hat b^\dagger_{nlm},\hat b^\dagger_{n'l'm'}]=0.
\end{equation}
The coefficients of the scalar field operator in eq. (\ref{quantizedphi})
must satisfy the Klein Gordon (KG) equation in a curved space time eq. (\ref{KG1}). 

Using the relations (\ref{conmutators}), one can construct the states
\begin{equation}\label{state}
|Q\rangle= |N_{nlm},N_{n'l'm'},N_{n''l''m''},...\rangle.
\end{equation}
These states are orthonormal and represent particle states, 
each composed of $\mathcal N$ scalar particles distributed in sets of 
$N_{nlm}$ particles of mass $\mu$, with angular momentum $\hbar l^i$ and 
azimuthal momentum $\hbar m^i$. The $n$ subindex labels the energy eigenstate. 
Then the expectation value of the energy momentum operator in 
(\ref{einsteineq}) can be calculated as 
\begin{equation}
\langle \hat T_{ab}\rangle\equiv \langle N_{nlm},N_{n'l'm'},\dots|\hat T_{ab}| N_{nlm},N_{n'l'm'},\dots\rangle\,. 
\end{equation}
The orthonomality of the quantum states ensures that this expectation value is 
given as a superposition of the expectation values of the energy-momentum for 
each state. Then (\ref{einsteineq}) is given by
\begin{equation}\label{einsteinequation}
G_{ab}=8\pi \sum_{nlm}c_{nlm}\langle  N_{nlm} 
|\hat T_{a b}|  N_{nlm}\rangle,
\end{equation}
where $c_{nlm}$ are normalization coefficients \cite{PhysRev.187.1767}.  
Therefore, in the case where more than one state is populated, the source of the
Einstein equations is equivalent to the superposition of many uncoupled
scalar fields. Each field generates its own stress energy tensor 
$\langle  N_{nlm} |\hat T_{a b}|  N_{nlm}\rangle$.


\section{The Z3 system in spherical symmetry, normal coordinates and
  regularization}
\label{the_Z3_system}

The line element of a generic spherically symmetric spacetime can be written as
\begin{equation}\label{line_element_ap}
ds^{2} = - \alpha^{2}dt^{2} + g_{rr}dr^{2} + r^{2}g_{\theta\theta}d\Omega^{2},
\end{equation}
where we made explicit the singular
factor $r^2$, such that the metric components are regular.
However, this change amounts to a transformation of the variables 
\begin{eqnarray}
\tilde{g}_{\theta\theta} &=& r^{2} g_{\theta\theta},\nonumber\\
\tilde{D}_{r\theta}{}^{\theta} &=& D_{r\theta}{}^{\theta} +
\frac{1}{r},\nonumber
\end{eqnarray}
where the quantities marked with \emph{tilde} are the variables typically used
in spherical symmetry.
In order to ensure the stability of the implementation, one has to deal with the factors $1/r$ in the fluxes and
$1/r^{2}$ in the sources. 

A regular system of evolution equations can be obtained, by ensuring a cross-cancellation between the
singular terms. 
We take advantage of the way the momentum constraint was built into the
system and redefine
the variable $Z_{r}$ in order to obtain the desired cross-cancellation,
\begin{eqnarray}
\tilde{Z_{r}} &=& Z_{r} +
\frac{1}{4r}\left(1-\frac{g_{rr}}{g_{\theta\theta}}\right).\nonumber
\end{eqnarray}
We can eliminate this way the singularities from the evolution variables
and the numerical errors caused by the geometrical factors in the fluxes and sources.
One can notice that the sources contain terms like $1/r$ times
other variables which are radial derivatives of the metric coefficients. But these
terms do not create problems at $r \rightarrow 0$, as the radial derivatives of
any differentiable function vanish at the origin. However, due to finite
differencing, we can not use a grid point at $r=0$. 

The final set of equations for the regularized Einstein-Klein-Gordon system in first
order form is:
\begin{eqnarray}
&& \partial_{t}g_{rr}          = -2\alpha g_{rr}K_{r}{}^{r},\nonumber\\
&& \partial_{t}g_{\theta\theta} = -2\alpha g_{\theta\theta}K_{\theta}{}^{\theta},\nonumber\\
&& \partial_{t}A_{r}         = -\partial_{r}[\alpha f trK],\nonumber\\
&& \partial_{t}D_{rr}{}^{r} = -\partial_{r}[\alpha K_{r}{}^{r}],\nonumber\\
&& \partial_{t}D_{r\theta}{}^{\theta} = -\partial_{r}[\alpha K_{\theta}{}^{\theta}],\nonumber\\
&& \partial_{t}Z_{r} = -\partial_{r}[2\alpha K_{\theta}{}^{\theta}] + \nonumber\\
&& + 2\alpha \left\{(K_{r}{}^{r} -
  K_{\theta}{}^{\theta})\left(D_{r\theta}{}^{\theta} + \frac{1}{r}\right) - \right.\nonumber\\
                    && - \left. K_{r}{}^{r}\left[Z_{r} +
                      \frac{1}{4r}\left(1-\frac{g_{rr}}{g_{\theta\theta}}\right)\right] +  A_{r}K_{\theta}{}^{\theta} + \right.\nonumber\\
&& + \left.  \frac{1}{4r}\frac{g_{rr}}{g_{\theta\theta}}(K_{\theta}{}^{\theta} -
  K_{r}{}^{r}) - 4 \pi \tau \right\},\nonumber\\
&& \partial_{t}K_{r}{}^{r} = -\partial_{r} \left[\alpha g^{rr} \left(A_{r} +
    \frac{2}{3}D_{r\theta}{}^{\theta} -\frac{4}{3}Z_{r}\right) \right] + \nonumber\\
                    && + \alpha \left\{(K_{r}{}^{r})^{2} +
                    \frac{2}{3}K_{\theta}{}^{\theta}(K_{r}{}^{r} -
                    K_{\theta}{}^{\theta}) - g^{rr}D_{rr}{}^{r} A_{r} + \right.\nonumber\\
                   && + \left. \frac{1}{3r}[g^{rr}(D_{rr}{}^{r} - A_{r} - 4Z_{r}) +
                     g^{\theta\theta}(D_{r\theta}{}^{\theta} - A_{r})]
                   \right. + \nonumber\\    
&& + \left. \frac{2}{3}g^{rr}\left[Z_{r} +
  \frac{1}{4r}\left(1-\frac{g_{rr}}{g_{\theta\theta}}\right)\right](2D_{rr}{}^{r} - 2D_{r\theta}{}^{\theta} - A_{r}) - \right. \nonumber\\
 && - \left. \frac{2}{3} g^{rr} \left( D_{r\theta}{}^{\theta} +
                          \frac{1}{r} \right)(D_{rr}{}^{r} - A_{r}) + \right.\nonumber\\
                     && + \left.  8\pi \left(\frac{\tau}{6} -
                        \frac{S_{r}{}^{r}}{2} + S_{\theta}{}^{\theta}\right) \right\},\nonumber\\
&& \partial_{t}K_{\theta}{}^{\theta} = - \partial_{r}\left[\alpha g^{rr}\left(-\frac{1}{3}D_{r\theta}{}^{\theta} +
    \frac{2}{3}Z_{r}\right)\right] + \nonumber\\ 
&& + \alpha \left\{\frac{1}{3}K_{\theta}{}^{\theta}(-K_{r}{}^{r} +
  4K_{\theta}{}^{\theta}) + \right.\nonumber\\
                        && + \left. \frac{1}{6r}[g^{rr}(A_{r} - 2D_{rr}{}^{r} -4Z_{r}) +
g^{\theta\theta}( A_{r} - 2D_{r\theta}{}^{\theta})] - \right.\nonumber\\
        && - \left. \frac{2}{3}g^{rr}\left[Z_{r} +
    \frac{1}{4r}\left(1-\frac{g_{rr}}{g_{\theta\theta}}\right)\right](D_{rr}{}^{r} - D_{r\theta}{}^{\theta} - 
                            2A_{r}) + \right.\nonumber\\
         && + \left. \frac{1}{3} g^{rr}\left(D_{r\theta}{}^{\theta} +
                 \frac{1}{r}\right)(D_{rr}{}^{r} - 4A_{r}) + \right.\nonumber\\
         && + \left. 8\pi \left(\frac{\tau}{6} -
             \frac{S_{r}{}^{r}}{2} + S_{\theta}{}^{\theta}\right)
                        \right\},\nonumber \\
 &&\partial_{t} \phi = \alpha \sqrt{g^{rr}} \phi_{t}, \nonumber\\
 &&\partial_{t} \phi_{r} = \partial_{r} [\alpha \sqrt{g^{rr}} \phi_{t}],\nonumber\\
 &&\partial_{t} \phi_{t} = \partial_{r} [\alpha \sqrt{g^{rr}} \phi_{r}] +
\alpha \sqrt{g^{rr}}[2(D_{r\theta}{}^{\theta} + 1/r)\phi_{r}+ \nonumber\\
&& + 2\sqrt{g_{rr}} K_{\theta}{}^{\theta} \phi_{t} - m^{2} g_{rr} \phi].\nonumber
\end{eqnarray}

The complex scalar field can decomposed as
\begin{eqnarray}
\phi &=& \phi^{R} - i\phi^{I},\nonumber\\
\bar{\phi} &=& \phi^{R} + i\phi^{I},\nonumber
\end{eqnarray}
where $\phi_{R}$ is the real part, $\phi_{I}$ the imaginary part and
$\bar{\phi}$ its complex conjugate.

The matter terms can be explicitly written in terms of the components
of the scalar field:
\begin{eqnarray}
\tau \ \ \ &=& \frac{1}{2}\{g^{rr}[(\phi^{I}_{t})^{2} + (\phi^{R}_{t})^{2}] + g^{rr}[(\phi^{I}_{r})^{2} +
(\phi^{R}_{r})^{2}] + \nonumber\\
& & + M^{2}[(\phi^{I})^{2} + (\phi^{R})^{2}]\} +
+\frac{1}{2}g^{rr}[(\psi_{t})^{2} +
(\psi_{r})^{2}], \nonumber\\
S_{r} \ \ &=& \sqrt{g^{rr}}(\phi^{I}_{t}\phi^{I}_{r} + \phi^{R}_{t}\phi^{R}_{r}) +
\sqrt{g^{rr}}\psi_{t}\psi_{r},\nonumber\\
S_{r}{}^{r} &=& \frac{1}{2}\{g^{rr}[(\phi^{I}_{t})^{2} + (\phi^{R}_{t})^{2}] + g^{rr}[(\phi^{I}_{r})^{2} +
(\phi^{R}_{r})^{2}] - \nonumber\\
& & - M^{2}[(\phi^{I})^{2} + (\phi^{R})^{2}]\} +\frac{1}{2}g^{rr}[(\psi_{t})^{2} +
(\psi_{r})^{2}],\nonumber\\
S_{\theta}{}^{\theta} &=& \frac{1}{2}\{g^{rr}[(\phi^{I}_{t})^{2} + (\phi^{R}_{t})^{2}] - g^{rr}[(\phi^{I}_{r})^{2} +
(\phi^{R}_{r})^{2}] - \nonumber\\
& & - M^{2}[(\phi^{I})^{2} + (\phi^{R})^{2}]\} +\frac{1}{2}g^{rr}[(\psi_{t})^{2} -
(\psi_{r})^{2}],\nonumber
\end{eqnarray}
where $\psi$ is the scalar field perturbation.

The charge density can be computed as
\begin{equation}\nonumber
\tilde{N} = \alpha J^{0} = \frac{1}{\sqrt{g_{rr}}}(\phi^{I}\phi^{R}_{t}-\phi^{R}\phi^{I}_{t}).
\end{equation}
The space volume integral of $\tilde{N}$ can be interpreted as the number of bosonic particles
\begin{equation}\nonumber
N = \int \sqrt{h} \ \tilde{N} \ dx^{3} = 4 \pi \int \ r^{2} \tilde{N}
\sqrt{g_{rr}} g_{\theta\theta} dr.
\end{equation}

We compute both the ADM and the Tolman masses in order to check our numerical evolutions.
The ADM mass is defined as
\begin{equation}\nonumber
M_{ADM} = \frac{1}{16\pi} \lim_{r \to \infty} \int g^{pq}[\partial_{q}g_{pk} -
\partial_{k}g_{pq}] N^{k} dS,
\end{equation}\nonumber
where $N^{r}=\sqrt{g^{rr}}\delta_{r}{}^{r}$ is the unit outward normal to the sphere.
In our coordinates, it can be translated into
\begin{equation}
M_{ADM} = - r^{2} \sqrt{g^{rr}} Dg_{r\theta}{}^{\theta}.
\end{equation}
The Tolman mass can be calculated as
\begin{eqnarray}
M_{Tol} &=& \int (T_{0}{}^{0} - T_{i}{}^{i})  \sqrt{-g} \ dx^{3} = \nonumber\\
&=& - 4 \pi r^{2}
\alpha \sqrt{g_{rr}} g_{\theta\theta}(\tau +
S_{r}{}^{r} + 2 S_{\theta}{}^{\theta}).\nonumber
\end{eqnarray}

%
%
\bibliography{./paper}
\bibliographystyle{apsrev}

\end{document}